\documentclass[prd,12pt,showpacs,floatfix,amsmath,amsfonts]{revtex4}

\usepackage{graphicx}
\usepackage{epstopdf}
\usepackage{bm}
\usepackage{multirow}

\DeclareFontFamily{OT1}{pzc}{}
\DeclareFontShape{OT1}{pzc}{m}{it}{<-> s * [1.200] pzcmi7t}{}
\DeclareMathAlphabet{\mathpzc}{OT1}{pzc}{m}{it}
\newcommand{\Eeff}{{\mathcal{E}}}
\newcommand{\Eexp}{{\mathpzc{E}}}

\newcommand\fdg{\mbox{$.\!\!^\circ$}}

\begin{document}

\title{Directional recoil rates for WIMP direct detection}

\author{Moqbil S. Alenazi}
\email{talianms@hotmail.com}
\author{Paolo Gondolo}
\email{paolo@physics.utah.edu}
\affiliation{Department of Physics, University of Utah, 115 S 1400 E Rm 201, Salt Lake City, Utah 84112-0830, USA}

\begin{abstract}
New techniques for the laboratory direct detection of dark matter weakly interacting massive particles (WIMPs) are sensitive to the recoil direction of the struck nuclei. We compute and compare the directional recoil rates ${dR}/{d\!\cos\theta}$ (where $\theta$ is the angle measured from a reference direction in the sky) for several WIMP velocity distributions including the standard dark halo and anisotropic models such as Sikivie's late-infall halo model and logarithmic-ellipsoidal models. Since some detectors may be unable to distinguish the beginning of the recoil track from its end (lack of head-tail discrimination), we introduce a ``folded'' directional recoil rate ${dR}/{d|\!\cos\theta|}$, where $|\!\cos\theta|$ does not distinguish the head from the tail of the track. We compute the CS$_2$ and CF$_4$ exposures required to distinguish a signal from an isotropic background noise, and find that ${dR}/{d|\!\cos\theta|}$ is effective for the standard dark halo and some but not all anisotropic models.
\end{abstract}

\pacs{95.35.+d}

\maketitle

\section{Introduction} 
\label{Introduction}

The nature of dark matter (DM) in the Universe is still one of the outstanding problems in astrophysics and cosmology. Numerous observations support the existence of DM. Examples are: big bang nucleosynthesis, cosmic microwave background data (WMAP3), supernova surveys, Galaxy surveys (SDSS, 2dF), and distance measurements with Cepheids (HST).
In the concordance cosmological model $\Lambda$CDM, the total density of the Universe has three contributions: matter, radiation, and a cosmological constant. The matter contribution can be further divided into the contribution of ordinary (baryonic) matter and the contribution of non-baryonic cold dark matter (CDM). Their density parameters, i.e.\ their densities in units of the critical density $\rho_{c}=1.053\times10^{-5} h^{2}({\rm GeV}/c^{2})\,{\rm cm}^{-3}$ (here $h$ is the Hubble constant in units of 100 km/s/Mpc), are:  $\Omega_{\rm b}=(0.02186 \pm 0.00068)\, h^{-2}$ and $\Omega_{\rm CDM}=(0.1105^{+0.0039}_{-0.0038})h^{-2}$ (from Ref.~\cite{LambdaCDM}). Thus, CDM constitutes $\sim 84\%$ of the matter in the Universe. 


CDM is found in clusters of galaxies and in individual spiral and elliptical galaxies. For example, stars in spiral galaxies are observed to move too fast around their galactic centers to be explained by the gravity of luminous matter alone. In particular, our Milky Way Galaxy also contains DM. Binney and Dehnen \cite{BD}, for example, show that the rotation curve of the Milky Way is nearly constant far beyond the Sun's location, implying the presence of DM in the Sun's neighborhood. This and similar studies give a density of DM near the Sun of $\rho = 0.3\,({\rm GeV}/c^{2})/{\rm cm}^{3}$.

The nature of CDM is still unknown. DM candidates for CDM are subatomic particles such as neutralinos, axions, Kaluza-Klein particles, and other WIMPs. WIMPs are hypothetical electrically-neutral stable particles with scattering cross-section off nucleons of the order of the weak interaction ($\sigma_{\rm p} \approx 10^{-44}\,{\rm cm}^{2}$) and mass in the range $10-1000$ GeV. Dark matter WIMPs arise for example as lightest supersymmetric particles (LSPs) in supersymmetric extensions of the Standard Model of particle physics. Kaluza-Klein particles arise in theories with more than four space-time dimensions, and share the same properties with WIMPs except for being somewhat heavier. In this paper, we refer to WIMPs but our considerations apply to Kaluza-Klein particles as well. 


Dark matter WIMPs near the Sun can reach the Earth and can scatter elastically off target nuclei in a detector, making the nuclei recoil. The energy, and recently the direction, of the recoiling nuclei can be measured experimentally. Extensive experimental efforts have been devoted to detect WIMPs directly (e.g., DAMA, CDMS, EDELWEISS, CRESST, DAMA/LIBRA, SuperCDMS, DRIFT, etc.).  There are two types of direct detection experiments: those that measure the recoil direction and those that do not. Examples of directional direct detectors are: DRIFT \cite{DRIFT1,DRIFT2,DRIFT3,DRIFT4,DRIFT5}, which uses a gas target in a time projection chamber and has run a prototype detector for a few years; NEWAGE \cite{NEWAGE1,NEWAGE2}, which uses a similar time projection chamber and is sensitive to spin-dependent WIMP-nucleus interactions; and detectors that use organic crystals such as stilbene \cite{Directional1,Directional2}. The other detectors previously listed are all non-directional.

A goal of directional WIMP detectors is to identify Galactic WIMPs by using the distribution of the nuclear recoil directions as a signature. We believe that an analysis of the WIMP-induced recoil directions can also allow the study of the structure and dynamics of the WIMP halo.

The idea of directional WIMP detection originated as early as 1988. Spergel \cite{Spergel} suggested that a WIMP signal could in principle be identified by means of the diurnal rotation of the `WIMP wind' direction due to the Earth's rotation (the `WIMP wind' is caused by the Solar System's rapid motion through the Galactic halo). The practical realization of directional WIMP detection was delayed by the difficulty of finding a suitable target material and an effective detection technique. In 1996, Martoff {\it et al.} \cite{Martoff} described a prototype direction-sensitive solid-state detector for WIMPs. A gaseous directional detector was studied by Martoff {\it et al.} \cite{DRIFT1} and is described in Snowden-Ifft, Martoff, and Burwell \cite{DRIFT2} (see also \cite{DRIFT3,DRIFT4,DRIFT5}). This detector, called DRIFT (Directional Recoil Identification From Tracks), uses a time projection chamber filled with a low pressure mixture of a target gas and an electronegative gas. The first stage of DRIFT (DRIFT I) had a 1 m$^3$ target (167 g of CS$_2$) and ran from 2001 to 2004 at the Boulby mine, North Yorkshire, England \cite{DRIFTI}. The current stage of DRIFT (DRIFT II) is an array of 1 m$^3$ modules and has been operational since 2005 \cite{DRIFTII}. A future stage has been envisaged (DRIFT III) that may have a target mass of up to 1 ton \cite{DRIFTIII}.

On the theoretical side, Copi, Heo, and Krauss \cite{CHK} and Copi and Krauss \cite{CK} examined the number of events required to distinguish a WIMP signal from an isotropic background. Gondolo \cite{Gon} obtained analytic expressions for a variety of directional recoil spectra by means of the Radon transform that relates the WIMP velocity distribution to the distribution of recoil momenta. Freese, Gondolo, and Newberg \cite{FGN} studied the possible directional detection of WIMPs belonging to the Sagittarius tidal stream, which may be showering DM onto the Solar System. Morgan, Green, and Spooner \cite{MGS}, Morgan and Green \cite{MorganGreen}, and Green and Morgan \cite{GreenMorgan} studied how the exposure required to directly detect a WIMP directional recoil signal depends on the capabilities of a directional detector. They also   examined statistical tests to distinguish a WIMP signal from an isotropic background and found that in detectors with head-tail discrimination (see below) of order ten events will be sufficient to distinguish a WIMP signal from an isotropic background for all of the halo models they considered. Host and Hansen \cite{HostHansen} investigated the possibility of measuring the velocity anisotropy of the Galactic dark matter halo in a direction-sensitive WIMP detector. They found that in excess of 10$^{5}$ events across all energies are needed to make a coarse measurement of the velocity anisotropy.

The goal of this paper is to study how different halo models affect the directional recoil rate ${dR}/{d\!\cos\theta}$, where $\theta$ is the angle between the nucleus recoil direction and a chosen reference direction in the sky. We compare the directional recoil rates for the different models. We repeat the same analysis for a ``folded'' directional recoil rate ${dR}/{d|\!\cos\theta|}$ that incorporates the inability of some detectors to distinguish the beginning of a recoil track from its end (head-tail discrimination). We compare each ${dR}/{d|\!\cos\theta|}$ to an isotropic background, to examine the possibility of discriminating a WIMP signal from background noise.

In Sec.~\ref{Directional Recoil Spectra}, we present a general discussion of the directional recoil spectra. There we give the expressions of various differential recoil rates that are useful to analyze and interpret WIMP direct detection experiments. In Sec.~\ref{Calculation of Recoil Momenta Distributions}, we describe two methods, numerical and analytical, for calculating the directional differential recoil rate ${dR}/{d\!\cos\theta}$ of recoiling target nuclei struck by WIMPs. The analytical method is applied to a Gaussian velocity distribution whose average velocity is aligned with the reference direction. The numerical method is more general and it can be used for any reference direction and any WIMP distribution. In Sec.~\ref{Results}, we present the results of applying the numerical method to various WIMP halo models, including streams of WIMPs, the standard dark halo, the Sikivie late-infall halo, and anisotropic models. In Sec.~\ref{Head-tail discrimination}, we address the difficulty of head-tail discrimination in WIMP direct detection experiments and present recoil distributions suitable for direct comparison with experiments lacking head-tail discrimination. Finally, we summarize our results in Sec.~\ref{Summary and Conclusions}.

\section{Directional Recoil Spectra}
\label{Directional Recoil Spectra}

In this section we give an expression for the directional recoil rate for interactions between WIMPs and target nuclei. In WIMP direct detection, the collision between the WIMP and the target nucleus is detected by measuring the energy of the recoiling nucleus. In directional detectors, one can also measure  its direction of recoil. 

\begin{figure}[ht]
\centering
\includegraphics[width=\textwidth]{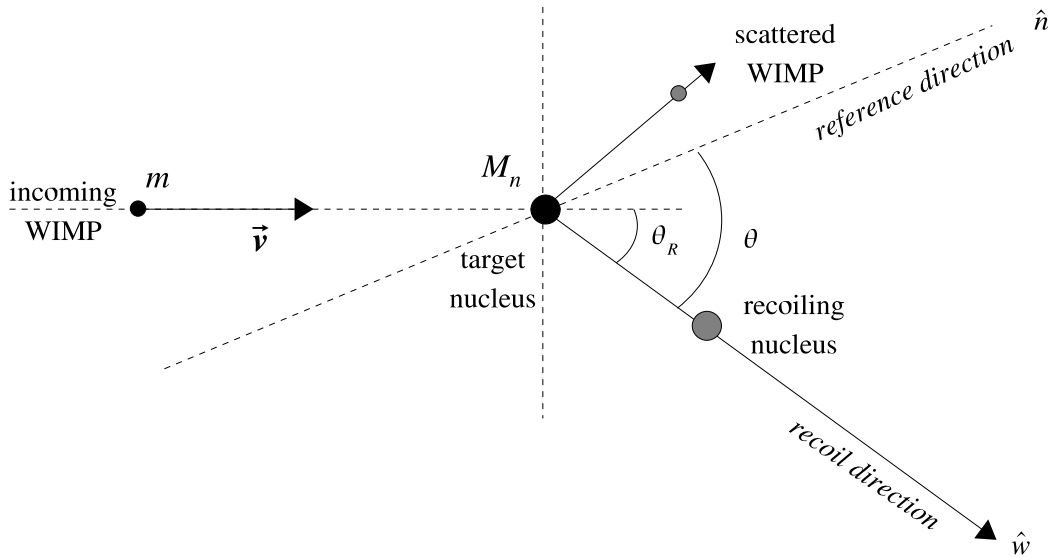}
\caption {Kinematics of a WIMP-nucleus elastic scattering.}
\label{fig3}
\end{figure}

Fig.~\ref{fig3} shows the kinematics of such a collision. The energy of the recoiling nucleus is given by (see e.g. Gascon \cite{Gas})
\begin{equation}
E = E_{\rm max} \cos^{2}\theta_{R} ,
\label{ER}
\end{equation}
where $\theta_R$ is the angle of the nuclear recoil relative to the initial WIMP direction (recoil angle), and
\begin{equation}
E_{\rm max} = \frac{2\mu_{n}^{2} v^{2}}{M_{n}}\,,
\end{equation}
is the maximum energy that the WIMP can transfer to the nucleus. Here $v$ is the speed of the incoming WIMP, $m$ is its mass, $M_{n}$ is the mass of the target nucleus, and $\mu_{n} = m\,M_{n}/(m + M_{n})$ is the reduced mass of the WIMP-nucleus system. 

In general, the differential recoil spectrum, i.e. the differential event rate per unit detector mass, is given by
\begin{equation}
\frac{d\,R}{d\,E} = \sum_{n} \frac{\rho}{2\,\mu_{n}^{2}\,m}\,C_{n}\,\sigma_{n}(E)\,\Eeff(E) \int_{v > w_n}\frac{f({\bf v})}{v}\,d^{3}\,v\,,
\label{dRdEgen}
\end{equation}
where the sum is over the nuclear species in the target, $C_{n}$ is the fraction of mass in species $n$,
\begin{equation}
w_{n} = c\,\sqrt{\frac{M_{n}\,E}{2\,\mu_{n}^{2}}}\,
\label{wn}
\end{equation}
 is the minimum WIMP speed required to transfer an amount of energy $E$ to the nucleus of mass $M_{n}$ in the detector (here $c$ is the speed of light), $\rho$ is the local WIMP density mentioned in the Introduction, $\Eeff(E)$ is the detection efficiency at recoil energy $E$, and $\sigma_{n}(E)$ is defined as
\begin{equation}
\sigma_{n}(E)=E_{\rm max}\, \frac{d\sigma}{dE}
\label{eq:sigma}
\end{equation}
 with $d\sigma/dE$ equal to the differential WIMP-nucleus scattering cross section.


For directional detectors, we need a differential rate not only in energy but also in direction. The three-dimensional recoil rate in spherical coordinates where the angles $\theta$ and $\phi$ refer to the direction of the nuclear recoil and the radial coordinate is the recoil energy $E$, is given by (see Ref.~\cite{Gon})
\begin{equation}
\frac{dR}{dE\,d\Omega} =  \sum_{n} \frac{\rho}{4\,\pi\,\mu_{n}^{2} \, m}\,C_{n}\,{\hat f}(w_n,{\bf \hat w})\,\sigma_{n}(E)\,\Eeff(E)\,,
\label{RecoilRate}
\end{equation}
where $d\Omega = d\phi\,d\cos\theta$  and ${\hat f}(w, {\bf\hat w})$ is the 3-dimensional Radon transform of the velocity distribution function $f({\bf v})$. The 3-dimensional Radon transform ${\hat f}(w, {\bf\hat w})$ of a function $f({\bf v})$ is defined to be the integral of $f({\bf v})$ on a plane orthogonal to the direction ${\bf\hat w}$ at a distance $w$ from the origin ${\bf v}=0$ \cite{Radon}. In formulas,
\begin{equation}
{\hat f}(w, {\bf\hat w}) = \int {\delta ({\bf v} \cdot {\bf\hat w} - w) \,f({\bf v}) \,d^{3} \,v} ,
\label{eq:Rint}
\end{equation}
where $\delta$ is the Dirac delta function and ${\bf\hat w}$ is the recoil direction. In this work we will specify the direction ${\bf \hat w}$ using a reference frame fixed in the sky in preference to a reference frame fixed with the laboratory.

\begin{figure}[bt]
\centering
\includegraphics[width=0.6\textwidth]{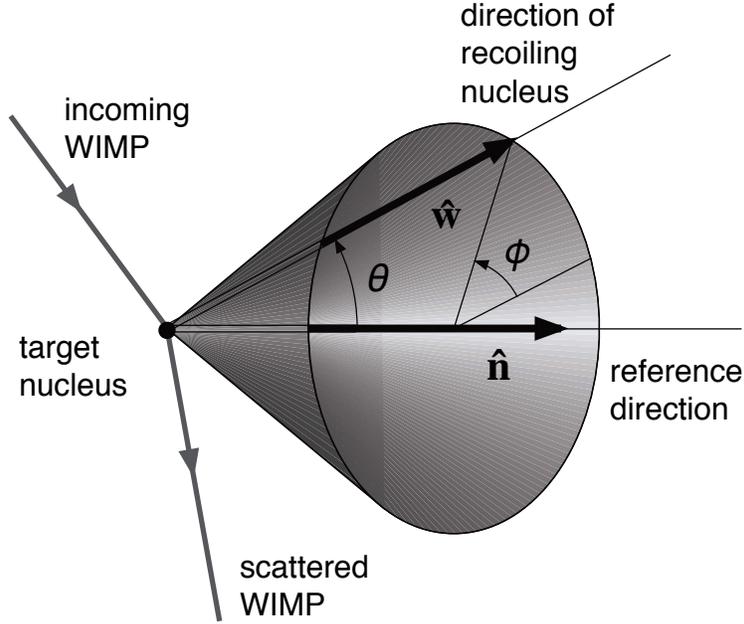}
\caption {The figure shows the reference direction ${\bf \hat n}$, pointing to a specific direction in the sky, and the angles $\theta$ and $\phi$ used in Sec.~\ref{Calculation of Recoil Momenta Distributions}.}
\label{whatnhat}
\end{figure}

Projections of the directional differential rate are also useful and have been used in the past. For example, one can measure recoil directions ${\bf\hat w}$ from a chosen reference direction ${\bf\hat n}$ as in Fig.~\ref{whatnhat}. If Eq.~(\ref{RecoilRate}) is integrated over the azimuthal angle $\phi$ and the energy $E$, one obtains
\begin{equation}
\frac{dR}{d\cos\theta} = \int \int{\frac{dR}{dE\,d\Omega}\,d\phi\,dE}\,,
\label{dRdcostheta}
\end{equation}
where $\theta$ is the angle between the reference direction ${\bf \hat n}$ and the recoil direction ${\bf \hat w}$. Eq.~(\ref{dRdcostheta}) is the directional differential recoil rate we study in this paper. It has been used in previous work \cite{NEWAGE1, Lehner} to compare WIMP velocity distributions and/or assess the advantages of directional detection methods.

The directional rate $dR/d\!\cos\theta$ requires a 3D read-out of the track direction. Although we are optimistic that one day a 3D read-out will be available, current experiments are limited to a 2D read-out in a plane fixed with the laboratory \cite{DRIFTII}. This plane precesses around the North-South terrestrial axis due to the rotation of the Earth. A differential rate $dR/d\phi$ appropriate for this situation has been introduced and studied in Refs.~\cite{MGS,MorganGreen,GreenMorgan}, to which we refer.

\begin{figure}[tb]
\centering
\includegraphics[width=\textwidth]{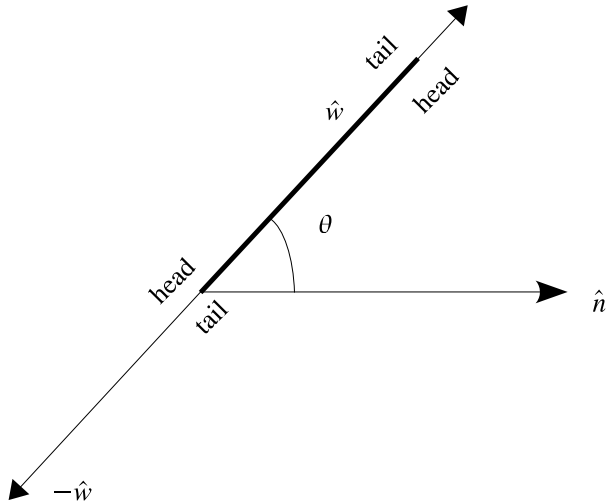}
\caption {The vectors $\hat{\bf w}$ and $-\hat{\bf w}$ share the same value of $|\!\cos\theta|$.}
\label{headtail}
\end{figure}

Besides the difficulty of a 3D read-out, current detectors may be unable to distinguish the beginning of the recoil track (the head) from the end of the track (the tail). This is called the difficulty of head-tail discrimination.
Because of this, it is useful to introduce the following ``folded'' directional recoil rate relevant to experiments that lack head-tail discrimination:
\begin{equation}
\frac{dR}{d|\!\cos\theta|} = \frac{dR(\cos\theta)}{d\cos\theta}+\frac{dR(-\cos\theta)}{d\cos\theta}\,.
\label{dRdcosabs}
\end{equation}
This rate is correctly normalized because the integral of both sides gives the total rate. As illustrated in Fig.~\ref{headtail}, $|\!\cos\theta|$ does not distinguish between ${\bf \hat w}$ and ${-\bf \hat w}$. Therefore, there is no need to know the track's head from its tail when using the ``folded'' directional recoil rate, Eq.~(\ref{dRdcosabs}). However, there may be a loss of information in doing so (see Section~\ref{Head-tail discrimination}).  

The differential WIMP-nucleus scattering cross section $\sigma_{n}(E)$ in Eq.~(\ref{eq:sigma}) can be split into two parts, one spin-independent (SI) and the other spin-dependent (SD):
\begin{equation}
\sigma_{n}(E) = \sigma_{n}^{SI}(E) + \sigma_{n}^{SD}(E)\,.
\label{sigman}
\end{equation}
Correspondingly, one can separate the spin-independent and spin-dependent contributions to the directional recoil rate $dR/{d\cos\theta}$ as
\begin{equation}
\frac{dR}{d\cos\theta} = \frac{dR^{\,SI}}{d\cos\theta} + \frac{dR^{\,SD}}{d\cos\theta}\,.
\label{dRdcosthetaTot}
\end{equation}
A similar separation can be defined for the ``folded'' directional recoil rate $dR/{d|\!\cos\theta|}$.

The rest of this section describes the expressions for the directional recoil rate for spin-independent and spin-dependent interactions.

\subsection{Spin-independent Directional Recoil Rates}
\label{Spin-independent Directional Recoil Rates}

In Eq.~(\ref{sigman}), the spin-independent part $\sigma_{n}^{SI}(E)$ can be written as
\begin{equation}
\sigma_{n}^{SI}(E) = \sigma_{0}\,\mathcal{F}_{n}(E)\,,
\label{sigmanSI}
\end{equation}
where $\sigma_{0}$ is the WIMP-nucleus scattering cross section and $\mathcal{F}_{n}(E)$ is a nuclear form factor which depends on the type of WIMP-nucleus interaction and on the mass and spin distributions within the nucleus. In cases where the nuclear form factor effects are negligible we have $\mathcal{F}_{n}(E) = 1$. In reality, the nuclear form factor may become important for specific detectors.

One can write
\begin{equation}
\sigma_{0} = \frac{\mu_{n}^{2}}{\pi}\,{|Z\,G_{s}^{\rm p} + (A - Z)\,G_{s}^{\rm n}|}^{2}\,,
\label{sigma0}
\end{equation}
where $Z$ is the number of protons in the nucleus, $A$ is the mass number of the nucleus, and $G_{s}^{\rm p}$\,($G_{s}^{\rm n}$) is the effective proton (neutron)-WIMP coupling. The WIMP-proton cross-section is
\begin{equation}
\sigma_{\rm p}\,=\,\frac{\mu_{\rm p}^{2}}{\pi}\,(G_{s}^{\rm p})^{2}\,,
\end{equation}
where
\begin{equation}
\mu_{\rm p} = \frac{\,m\,m_{\rm p}}{m+m_{\rm p}}\,
\end{equation}
is the WIMP-proton reduced mass.
Assuming, as it is approximately the case for neutralino dark matter, that
\begin{equation}
G_{s}^{\rm p}=G_{s}^{\rm n}\,,
\end{equation}
we have
\begin{equation}
\sigma_{0} = \frac{\mu_{n}^{2}}{\mu_{\rm p}^{2}}\,A^{2}\,\sigma_{\rm p}\,.
\label{sigma01}
\end{equation}
In this case, the recoil rate, Eq.~(\ref{RecoilRate}), takes the form
\begin{equation}
{\frac{dR^{\,SI}}{dE\,d\Omega}} = \frac{\rho\,\sigma_{\rm p}}{4\,\pi\,\mu_{\rm p}^{2}\,m}\,\sum_{n} C_{n}\,A_{n}^{2}\,{\hat f}(w_{n},{\bf \hat w})\,\mathcal{F}_{n}(E) \,\Eeff(E)\,.
\label{totrate}
\end{equation}
We can define an effective spin-independent recoil momentum distribution, ${\hat f}_{\rm eff}^{SI}(E,{\bf \hat w})$, as the average over all masses
\begin{equation}
{\hat f}_{\rm eff}^{SI}(E,{\bf \hat w}) = \sum_{n} C_{n}\,A_{n}^{2}\,{\hat f}(w_{n},{\bf \hat w})\,\mathcal{F}_{n}(E) \,\Eeff(E)\,.
\label{fhateff}
\end{equation}
The rate of detection of WIMPs then reads
\begin{equation}
{\frac{dR^{\,SI}}{dE\,d\Omega}} = \frac{\rho\,\sigma_{\rm p}}{4\,\pi\,\mu_{\rm p}^{2}\,m}\,{\hat f}_{\rm eff}^{SI}(E,{\bf \hat w})\,.
\label{totrate2}
\end{equation}
We can also write
\begin{equation}
{\frac{dR^{\,SI}}{d\!\cos\theta}} = \frac{\rho\,\sigma_{\rm p}}{4\,\pi\,\mu_{\rm p}^{2}\,m}\,\int\int{{\hat f_{\rm eff}}^{SI}(E,{\bf \hat w})\,d\phi}\,dE.
\label{dRbydEdcosfhat}
\end{equation}

For example, for a CS$_{2}$ target as in DRIFT,
\begin{equation}
C_{\rm S} = \frac{2\,M_{\rm S}}{2\,M_{\rm S}+M_{\rm C}}\,,
\end{equation}
\begin{equation}
C_{\rm C} = \frac{M_{\rm C}}{2\,M_{\rm S}+M_{\rm C}}\,,
\end{equation}
and ${\hat f}_{\rm eff}^{SI}(E,{\bf \hat w})$ is given explicitly by
\begin{equation}
{\hat f}_{\rm eff}^{SI}(E,{\bf \hat w}) = \frac{2\,M_{\rm S}\,A_{\rm S}^{2}\,{\hat f}_{\rm S}\,\mathcal{F}_{\rm S}+M_{\rm C}\,A_{\rm C}^{2}\,{\hat f}_{\rm C}\, \mathcal{F}_{\rm C}}{2\,M_{\rm S}+M_{\rm C}}\,\Eeff.
\label{fhatCfhatS}
\end{equation}
Here ${\hat f}_{n}={\hat f}(w_{n},{\bf \hat w})$. (Notice that the symbol $C$ (italic) denotes the fraction of mass while the symbol/subscript symbol ${\rm C}$ (roman) denotes the carbon nucleus).

Using common units and magnitudes, the spin-independent directional detection rate of WIMPs is
\begin{equation}
{\frac{dR^{\,SI}}{dE\,d\Omega}} = 1.306\times10^{-3}\,\frac{\rm events}{\hbox{\rm kg-day-keV-sr}}\,\frac{\rho_{0.3}\,\sigma_{44}}{4\,\pi\,\mu_{\rm p}^{2}\,m}\,{\hat f}_{\rm eff}^{SI}(E, {\bf\hat w})\,,
\label{ftotrate}
\end{equation}
where $\rho_{0.3}$ is the DM density in the solar neighborhood in units of $0.3\,({\rm GeV/c^{2}})/{\rm cm^{3}}$, $\sigma_{44}$ is the proton cross-section in units of $10^{-44}{\rm cm^{2}}$. $\mu_{\rm p}$ and $m$ are in ${\rm GeV/c^{2}}$, and ${\hat f}_{\rm eff}^{SI}$ is in $({\rm km/s})^{-1}$.

\subsection{Spin-dependent Directional Recoil Rates}
\label{Spin-dependent Directional Recoil Rates}

In Eq.~(\ref{sigman}), the spin-dependent part $\sigma_{n}^{SD}(E)$ can be written as
\begin{equation}
\sigma_{n}^{SD}(E) = \frac{32\,\mu_{n}^{2}\,G_{F}^{2}}{(2\,J_{n}+1)\hbar^4}\,[a_{\rm p}^{2}\,S_{{\rm pp}}^{n}(E) + a_{\rm n}^{2}\,S_{{\rm nn}}^{n}(E) + a_{\rm p}\,a_{\rm n}\,S_{{\rm pn}}^{n}(E)]\,.
\label{sigmanSD}
\end{equation}
Here $\hbar$ is the reduced Planck constant, $G_{F}$ is Fermi coupling constant ($G_{F}/(\hbar c)^3=1.16637 \times 10^{-5}$\,GeV$^{-2}$), $J_{n}$ is the nucleus total angular momentum in units of $\hbar$, $a_{\rm p}\,(a_{\rm n})$ is the effective axial coupling of WIMP and proton (neutron) in units of $2\sqrt{2}\,G_{F}/\hbar^2$ \cite{SavGonFre}. (Notice that the subscript/superscript symbol $n$ (italic) denotes the nucleus while the subscript symbol ${\rm n}$ (roman) denotes the neutron). In Eq.~(\ref{sigmanSD}), the dimensionless functions $S_{{\rm pp}}^{n}(E)$, $S_{{\rm nn}}^{n}(E)$, and $S_{{\rm pn}}^{n}(E)$ play the same role as the nuclear form factor $\mathcal{F}_{n}(E)$ in the spin-independent case. They are given by 
\begin{eqnarray}
S_{{\rm pp}}^{n} & = & S_{00}^{n}+S_{11}^{n}+S_{01}^{n}\,,
\\
 S_{{\rm nn}}^{n} & = & S_{00}^{n}+S_{11}^{n}-S_{01}^{n}\,,
\\
S_{{\rm pn}}^{n} & = & 2(S_{00}^{n}-S_{11}^{n})\,,
\label{Ss}
\end{eqnarray}
where $S_{00}^{n}$, $S_{11}^{n}$, $S_{01}^{n}$ are the nuclear spin structure functions defined in \cite{Engel}. When the nuclear spin is approximated by the spin of the odd nucleon only, one finds
\begin{equation}
S_{{\rm pp}}^{n} = \frac{\lambda_{n}^{2}\,J_{n}(J_{n}+1)(2J_{n}+1)}{\pi}\,,\,\,S_{{\rm nn}}^{n} = 0\,,\,\,S_{{\rm pn}}^{n} = 0\,,
\label{Spp}
\end{equation}
for a proton-odd nucleus, and
\begin{equation}
S_{{\rm pp}}^{n} = 0\,\,,\,S_{{\rm nn}}^{n} = \frac{\lambda_{n}^{2}\,J_{n}(J_{n}+1)(2J_{n}+1)}{\pi}\,,\,\,S_{{\rm pn}}^{n} = 0\,,
\end{equation}
for a neutron-odd nucleus. Here $\lambda_{n}$ is conventionally defined through the relation $<\!n|{\bf S}|n\!> = \lambda_{n}<\!n|{\bf J}|n\!>$, where $|n\!>$ is the nuclear state, ${\bf S}$ is the spin, ${\bf J}$ is the total angular momentum. Tables of $\lambda_{n}^{2}J_{n}(J_{n}+1)$ values for several nuclei can be found in \cite{EllisFlores} and \cite{LewinSmith}.

The spin-dependent cross-section off a proton is
\begin{equation}
\sigma_{\rm p}^{SD} = \frac{24\,\mu_{\rm p}^{2}\,G_{F}^{2}}{\pi\hbar^4}\,a_{\rm p}^{2}\,,
\end{equation}
and that off a neutron is
\begin{equation}
\sigma_{\rm n}^{SD} = \frac{24\,\mu_{\rm n}^{2}\,G_{F}^{2}}{\pi\hbar^4}\,a_{\rm n}^{2}\,.
\end{equation}

In case the target is a combination of different nuclei, we write
\begin{equation}
{\frac{dR^{\,SD}}{dE\,d\Omega}} = \frac{6\,\rho\,G_{F}^{2}}{\pi^{2}\,m\,\hbar^4}\,\left[a_{\rm p}^{2}\,{\hat f}_{{\rm eff},{\rm pp}}^{SD}(E,{\bf \hat w})+a_{\rm n}^{2}\,{\hat f}_{{\rm eff},{\rm nn}}^{SD}(E,{\bf \hat w})+a_{\rm p}a_{\rm n}\,{\hat f}_{{\rm eff},{\rm pn}}^{SD}(E,{\bf \hat w})\right]\,,
\label{totrateSD}
\end{equation}
where we define the effective spin-dependent recoil momentum distributions
\begin{equation}
{\hat f}_{{\rm eff},{\rm pp}}^{SD}(E,{\bf \hat w}) = \sum_{n}\,\frac{4\,\pi}{3(2J_{n}+1)}\,C_{n}\,S_{{\rm pp}}^{n}(E)\,{\hat f}(w_{n},{\bf \hat w})\,\Eeff(E),
\label{fhatppSD}
\end{equation}
and similarly for ${\hat f}_{{\rm eff},{\rm nn}}^{SD}(E,{\bf \hat w})$ and ${\hat f}_{{\rm eff},{\rm pn}}^{SD}(E,{\bf \hat w})$. The normalization of ${\hat f}_{{\rm eff},{\rm pp}}^{SD}(E,{\bf \hat w})$, ${\hat f}_{{\rm eff},{\rm nn}}^{SD}(E,{\bf \hat w})$, and ${\hat f}_{{\rm eff},{\rm pn}}^{SD}(E,{\bf \hat w})$ has been chosen so that ${\hat f}_{{\rm eff},{\rm pp}}^{SD}(E,{\bf \hat w})={\hat f}(w_{\rm p},{\bf \hat w})$ when the target is a proton and ${\hat f}_{{\rm eff},{\rm nn}}^{SD}(E,{\bf \hat w})={\hat f}(w_{\rm n},{\bf \hat w})$ when it is a neutron.

For the CF$_{4}$ target used in the NEWAGE detector \cite{NEWAGE1,NEWAGE2}, and in the proton-odd approximation, Eq.~(\ref{totrateSD}) takes the form
\begin{equation}
{\frac{dR^{\,SD}}{dE\,d\Omega}} = \frac{\rho\,\sigma_{\rm p}^{SD}}{4\,\pi\,\mu_{\rm p}^{2}\,m}\,{\hat f}_{{\rm eff},{\rm pp}}^{SD}(E,{\bf \hat w})\,.
\label{totrateSDCF4}
\end{equation}
In this case, the C nucleus has no spin (thus $\lambda_{\rm C}=0$), while the F nucleus has spin $\frac{1}{2}$. In the proton-odd approximation, ${\lambda_{\rm F}^{2}}J_{\rm F}(J_{\rm F}+1)=0.647$ (see Table 1 in \cite{EllisFlores}). Thus the effective spin-dependent recoil momentum distribution, Eq.~(\ref{fhatppSD}), reads
\begin{equation}
{\hat f}_{{\rm eff},{\rm pp}}^{SD}(\rm {CF_{4}}) = \frac{4}{3}\,\,0.647\,{\it C}_{F}\,{\it {\hat f}}_{F}\,\Eeff({\it E}).
\label{fhatppSDCF4}
\end{equation}
The fraction of mass $C_{\rm F}$ is given by
\begin{equation}
C_{\rm F} = \frac{4\,M_{\rm F}}{4\,M_{\rm F}+M_{\rm C}}\,.
\end{equation}
Using common units and magnitudes, the spin-dependent directional detection rate of WIMPs (for the CF$_{4}$ target used in the NEWAGE detector) is
\begin{equation}
{\frac{dR^{\,SD}}{dE\,d\Omega}} = 1.306\times10^{-3}\,\frac{\rm events}{\hbox{\rm kg-day-keV-sr}}\,\frac{\rho_{0.3}\,\sigma_{{\rm p},44}^{SD}}{4\,\pi\,\mu_{\rm p}^{2}\,m}\,{\hat f}_{{\rm eff},{\rm pp}}^{SD}(E, {\bf\hat w})\,.
\label{ftotrateSD}
\end{equation}

\section{Calculation of $\bm{{dR}/{d\!\cos\theta}}$}
\label{Calculation of Recoil Momenta Distributions}

The analysis of directional spectra in WIMP direct detection can be carried out by computing the directional differential recoil rate ${dR}/{d\!\cos\theta}$ as a function of the angle $\theta$ between the nuclei's recoil directions ${\bf \hat w}$ and a reference direction ${\bf \hat n}$. In this section we describe two methods for calculating the directional recoil rate ${dR}/{d\!\cos\theta}$. The first method is numerical and can be used for any WIMP velocity distribution and for any reference direction. The second method is analytic and is restricted to Gaussian distributions and to reference directions ${\bf \hat n}$ in the same direction as the WIMP average velocity ${\bf V}$. Results of applying these methods to various dark halo models are given in Sec.~\ref{Results}.

Before describing the two methods, we recall the expression of the recoil momentum function ${\hat f}(w, {\bf\hat w})$ for a Maxwellian velocity distribution. In the rest frame of the detector, the WIMP velocity distribution is given by
\begin{equation}
f({\bf v}) = \frac{1}{(2 \pi \sigma_{v}^2)^{3/2}} \exp\!{\left(- {\frac{\left| {\bf v} - {\bf V} \right|^2}{2 \sigma_{v}^2}}\right)} ,
\label{fMax}
\end{equation}
where ${\bf v}$ is the velocity of a WIMP,  $\sigma_{v}$ is the velocity dispersion (not to be confused with the WIMP-nucleus cross section) and ${\bf V}$ is the average velocity of the WIMPs with respect to the detector. 
The recoil momentum spectrum for nucleus $n$ in the laboratory frame is \cite{Gon}
\begin{equation}
{\hat f}_{n}(w, {\bf\hat w}) = \frac{1}{(2 \pi \sigma_{v}^2)^{1/2}} \exp\!{\left(- {\frac{[w_{n}-{{\bf\hat w}\cdot{\bf V}}]^{2}}{2 \sigma_{v}^2}}\right)}\,.
\label{eq:RC}
\end{equation}
In principle,
\begin{equation}
{\bf V} = {\bf V}({\rm W},{\rm G}) - {\bf V}({\rm S},{\rm G}) - {\bf V}({\rm E},{\rm S}) - {\bf V}({\rm lab},{\rm E}),
\end{equation}
where ${\bf V}({\rm W},{\rm G})$ is the average velocity of the WIMPs relative to the Galactic rest frame (zero in the standard halo model), ${\bf V}({\rm S},{\rm G})$ is the velocity of the Sun relative to the Galactic rest frame (of order 200 km/s), ${\bf V}({\rm E},{\rm S})$ is the velocity of the Earth relative to the Sun (of order 30 km/s), and ${\bf V}({\rm lab},{\rm E})$ is the velocity of the detector in the laboratory relative to the center of mass of the Earth (of order 0.3 km/s). As the Earth rotates and orbits the Sun, two signal modulations (annual and diurnal) are expected as a result of the relative motions ${\bf V}({\rm E},{\rm S})$ and ${\bf V}({\rm lab},{\rm E})$. In this work we neglect ${\bf V}({\rm E},{\rm S})$ and ${\bf V}({\rm lab},{\rm E})$ -- and the corresponding annual and diurnal modulations -- and use the velocity distribution in the frame of the Sun. We also neglect focusing effects due to the gravitational field of the Sun (of order 1 km/s, see \cite{AG}).

In the following, we assume that the nuclear form factors are $\mathcal{F}_{n}(E)=1$. In the case of a detector with threshold energy $E_{\rm thr}$, we model the detection efficiency at recoil energy $E$ as
\begin{equation}
\Eeff(E) = \left\{
\begin{array}{l l}
\displaystyle
 1 \, , & \quad \mbox{if $E$ $>E_{\rm thr}$\,,}\\
  0, & \quad \mbox{if $E$ $<E_{\rm thr}$\,.}\\ \end{array} \right.
\end{equation}

\subsection{Numerical $\bm{{dR}/{d\!\cos\theta}}$ - general case}
\label{numerical for general cases}

In general directions and for non-Gaussian distributions, the integration of the recoil rate over the energy $E$ and the azimuthal angle $\phi$ in  Eq.~(\ref{dRdcostheta}) cannot be done analytically. However, one can calculate the recoil distribution numerically. 

We do the integration over the azimuthal angle $\phi$ using a Riemann sum. For the integration over the energy $E$, it is difficult to use a Riemann sum because of narrow peaks in ${dR}/{d\!\cos\theta}$ as a function of $\cos\theta$ when streams of dark matter are present. Therefore, we do the integration over $E$ by means of the fifth-order Cash-Karp Runge-Kutta method with adaptive stepsize control, as described in Ref.~\cite{NR}.
To apply this method, we write the differential recoil rate in the form of an ordinary differential equation
\begin{equation}
\frac{d}{dE}\left(\frac{dR}{d\cos\theta}\right) = 1.306\times10^{-3}\,\frac{\rho_{\rm 0.3}\,\sigma_{\rm 44}}{4\,\pi\,\mu_{\rm p}^{2}\,m}\,\sum_{i=1}^{N_{\rm \phi}}\,{\hat f}_{\rm eff}(E,{\bf \hat w}_i)\,\Delta\phi\,,
\label{dRdcostheRK}
\end{equation}
where $\Delta\phi = {2\,\pi}/{N_{\rm \phi}}$, and the unit vectors ${\bf \hat w}_i$, which depend on $\phi_{i}$, are specified in Eq.~(\ref{wzhatn}) below.
We take $N_{\rm \phi}={\rm 100}$ with initial condition $(dR/d\cos\theta)_{E=0}=0$; we require an accuracy of $10^{-8}$ and choose a scaling factor  
\begin{equation}
y_{i}^{\rm scal} = {\left|\, y_{i}\, \strut \right|} + {\left|h_{\rm RK}\,{\left(\frac{dy}{dE}\right)}_{i} \strut \right|}\,,
\label{yscal}
\end{equation}
where $h_{\rm RK}$ is the value of the stepsize, $y_i = dR/d\!\cos\theta$ at $\phi=\phi_i$, and $({dy}/{dE})_i$ is the partial derivative of $dR/d\!\cos\theta$ with respect to $E$ at $\phi=\phi_i$.

An expression for ${\bf \hat w}_i$ is obtained as follows. At a fixed value of the angle $\theta$ between the reference direction ${\bf \hat n}$ and the recoil direction ${\bf \hat w}$, the possible directions of the target nucleus recoils lie on a cone of opening angle $2\theta$, as shown in Fig.~\ref{whatnhat}. Since $\theta$ is constant, it is convenient to specify ${\bf \hat w}$ in terms of the polar angle $\theta$ and the azimuthal angle $\phi$ reckoned with respect to the ${\bf \hat n}$ axis. 

We want to find the Cartesian components of ${\bf \hat w}_i$ in an arbitrarily given reference frame $x$, $y$, $z$. For this purpose, we introduce an auxiliary Cartesian coordinate system $x'$, $y'$, $z'$, with $z'$ aligned with ${\bf \hat n}$. Since $\theta$ and $\phi$ are reckoned from $z'$, the vector ${\bf \hat w}_i$, which is ${\bf\hat w}$ at $\phi=\phi_i$, can be written in terms of the new unit vectors ${\bf \hat e'}_{1}$, ${\bf \hat e'}_{2}$, ${\bf \hat e'}_{3}$ as:
\begin{equation}
{\bf \hat w}_i = \sin\theta\,\cos\phi_{i}\,{\bf \hat e'}_{1} + \sin\theta\,\sin\phi_{i}\,{\bf \hat e'}_{2} + \cos\theta\,{\bf \hat e'}_{3}\,.
\label{wihat}
\end{equation}

Our task is now to express the new basis vectors ${\bf \hat e'}_{1}$, ${\bf \hat e'}_{2}$, ${\bf \hat e'}_{3}$ in terms of the original basis vectors ${\bf \hat e}_{1}$, ${\bf \hat e}_{2}$, ${\bf \hat e}_{3}$. Once this is achieved, Eq.~(\ref{wihat}) will give ${\bf \hat w}_i$ in terms of ${\bf \hat e}_{1}$, ${\bf \hat e}_{2}$, ${\bf \hat e}_{3}$ and one can read off  its Cartesian components.

The transformation matrix from ${\bf \hat e}_{1}$, ${\bf \hat e}_{2}$, ${\bf \hat e}_{3}$ to ${\bf \hat e'}_{1}$, ${\bf \hat e'}_{2}$, ${\bf \hat e'}_{3}$ can be found using Euler angles. We first rotate the ${\bf\hat e}_{i}$ axes about the $z$ axis counterclockwise through an angle $\alpha+\frac{\pi}{2}$. Then we rotate the resulting axes about the new $x$ axis counterclockwise through an angle $\beta$. We find 
\begin{eqnarray}
{\bf \hat e'}_{1} & = & - \sin\alpha\,{\bf \hat e}_{1} + \cos\alpha\,{\bf \hat e}_{2}\,,
\label{ed1}
\\
{\bf \hat e'}_{2} & = & - \cos\beta\,\cos\alpha\,{\bf \hat e}_{1} - \cos\beta\,\sin\alpha\,{\bf \hat e}_{2} + \sin\beta\,{\bf \hat e}_{3}\,,
\label{ed2}
\\
{\bf \hat e'}_{3} & = & \sin\beta\,\cos\alpha\,{\bf \hat e}_{1} + \sin\beta\,\sin\alpha\,{\bf \hat e}_{2} + \cos\beta\,{\bf \hat e}_{3}\,.
\label{ed3}
\end{eqnarray}
Identifying ${\bf \hat n}$ with ${\bf \hat e}_{3}$ in Eqs.~(\ref{ed1}\,-\,\ref{ed3}) shows that $\alpha$ and $\beta$ are the spherical coordinates of ${\bf \hat n}$:
\begin{eqnarray}
\alpha & = & \tan^{-1} \left(\frac{{{\rm n}_{\rm y}}}{{\rm n}_{\rm x}}\right)\,,
\\
\beta & = & \cos^{-1} \left({\rm n}_{\rm z}\right)\,.
\end{eqnarray}

Inserting these relations into Eq.~(\ref{wihat}), we obtain
\begin{eqnarray}
{\bf \hat w}_{i} & = & (- \sin\theta\,\cos\phi_{i}\,\sin\alpha - \cos\beta\,\cos\alpha\,\sin\theta\,\sin\phi_{i}\,+\sin\beta\,\cos\alpha\,\cos\theta)\,{\bf \hat e}_{1}
\nonumber \\
 & + & (\sin\theta\,\cos\phi_{i}\,\cos\alpha - \sin\theta\,\sin\phi_{i}\,\cos\beta\,\sin\alpha + \sin\beta\,\sin\alpha\,\cos\theta)\,{\bf \hat e}_{2}
\nonumber \\
 & + & (\sin\theta\,\sin\phi_{i}\,\sin\beta + \cos\beta\,\cos\theta)\,{\bf \hat e}_{3}\,.
\label{wzhatn}
\end{eqnarray}

\subsection{Analytic $\bm{{dR}/{d\!\cos\theta}}$ - special case}
\label{Analytic for fixed reference direction}

We were able to derive an analytic expression for ${dR}/{d\!\cos\theta}$ when the WIMP velocity distribution is Maxwellian and the reference direction ${\bf \hat n}$ is aligned with the WIMP average velocity ${\bf V}$. Notice that this does not mean that $\theta=\theta_{R}$, the recoil angle (see Fig.~\ref{fig3}), because $\theta_R$ is measured from the velocity of an individual WIMP while $\theta$ is measured from the {\it average} velocity of all WIMPs. Here we assume a spin-independent case and a zero-threshold detector.

We wish to compute
\begin{equation}
\frac{dR}{d\cos\theta} =  \frac{\rho\,\sigma_{\rm p}}{4\,\pi\,\mu_{\rm p}^{2}\,m}\, \int \int {\hat f}_{\rm eff}(E,{\bf \hat w})\,dE\,d\phi\, .
\end{equation}
Using Eq.~(\ref{fhateff}), 
\begin{equation}
\frac{dR}{d\cos\theta} = \frac{\rho\,\sigma_{\rm p}}{4\,\pi\,\mu_{\rm p}^{2}\,m}\, \sum_{n} C_n A_n^2 I_n ,
\label{eq:40}
\end{equation}
where
\begin{equation}
I_n =  \int_{0}^{\infty} \int_{0}^{2\pi} \hat f_{n}(w, {\bf \hat w})\,dE\,d\phi\,.
\label{eq:In}
\end{equation}
For a Maxellian velocity distribution,
\begin{equation}
I_n =  \frac{1}{(2 \pi \sigma_{v}^2)^{1/2}} \int_{0}^{\infty} \int_{0}^{2\pi} \exp\!{\left(- {\frac{[w_{n}-{{\bf\hat w}\cdot{\bf V}}]^{2}}{2 \sigma_{v}^2}}\right)}\,dE\,d\phi\,.
\label{I}
\end{equation}
Since ${\bf \hat w} \cdot {\bf \hat n} = \cos\theta$ and we assume here that ${\bf V} = V {\bf \hat n}$, we have
\begin{equation}
{\bf \hat w} \cdot {\bf V} = V\cos\theta\,.
\label{Vcos}
\end{equation}
{}From Eq.~(\ref{wn}),
\begin{equation}
dE = \frac{4\,\mu_{n}^{\rm 2}}{M_{n}\,c^{\rm 2}}\,w_{n}\,dw_n\,.
\label{dE}
\end{equation}
Substituting Eqs.~(\ref{Vcos}) and (\ref{dE}) into Eq.~(\ref{I}), we find
\begin{eqnarray}
I_n & = & \frac{8\,\pi\,\mu_{n}^{\rm 2}}{M_{n}\,c^{\rm 2}\,(2\,\pi\,\sigma_{v}^{2})^{\rm 1/2}}\,\int_{0}^{\infty}\exp\!{\left(- \frac{[w_{n} - V\cos\theta]^{\rm 2}}{2\,\sigma_{v}^{2}}\right)}\,w_{n}\,dw_n\,,
\\
& = & \frac{8\,\pi\,\mu_{n}^{2}\,\sigma_{v}}{M_{n}\,c^{2}\,(2\,\pi)^{1/2}}\,{\left\{{\rm exp}\left({-\frac{V^{2}\!\cos^{2}\theta}{2\,\sigma_{v}^{2}}}\right) + \frac{V\cos\theta}{\sigma_{v}\,\sqrt{2}}\,\sqrt{\pi}\,{\left[1+{\rm erf}{\left(\frac{V\cos\theta}{\sigma_{v}\,\sqrt{2}}\right)}\right]}\right\}}\,,
\label{I1}
\end{eqnarray}
where ${\rm erf}(x)$ is the error function. Inserting Eq.~(\ref{I1}) into Eq.~(\ref{eq:40}) gives
\begin{equation}
\frac{dR}{d\cos\theta} =  \sum_{n} \frac{2\rho\sigma_{\rm p}C_n A_n^2\mu_{n}^{2}\sigma_{v}}{\sqrt{2\pi}M_{n}c^{2}m\mu_{\rm p}^{2}} {\left\{{\rm exp}\left({-\frac{V^{2}\!\cos^{2}\theta}{2\sigma_{v}^{2}}}\right) + \frac{V\cos\theta}{\sigma_{v}\sqrt{2}}\sqrt{\pi}{\left[1+{\rm erf}{\left(\frac{V\cos\theta}{\sigma_{v}\sqrt{2}}\right)}\right]}\right\}}
\end{equation}
or
\begin{eqnarray}
\frac{dR}{d\cos\theta} & = & 2612 \, \frac{\rm events}{\hbox{\rm kg-day}} \, \sum_{n} \frac{\rho_{0.3}\sigma_{44}C_n A_n^2\mu_{n}^{2}\sigma_{v}}{\sqrt{2\pi}M_{n}c^{2}m\mu_{\rm p}^{2}} \times \nonumber \\ && {\left\{{\rm exp}\left({-\frac{V^{2}\!\cos^{2}\theta}{2\sigma_{v}^{2}}}\right) + \frac{V\cos\theta}{\sigma_{v}\sqrt{2}}\sqrt{\pi}{\left[1+{\rm erf}{\left(\frac{V\cos\theta}{\sigma_{v}\sqrt{2}}\right)}\right]}\right\}}.
\end{eqnarray}

This analytic formula agrees with the numerical method, Eq.~(\ref{dRdcostheRK}), in the cases in common.

\section{Results}
\label{Results}

In this section we present the results of applying the numerical method, Eq.~(\ref{dRdcostheRK}), to various WIMP halo models, including streams of WIMPs, the standard dark halo, the Sikivie late-infall halo (SLI streams), and anisotropic logarithmic-ellipsoidal models. For comparison's sake, in the last three cases (the standard dark halo, SLI streams, and the anisotropic models), we fix the reference direction ${\bf \hat n}$ to be the direction of Galactic rotation. For all these models, we use CS$_{2}$ molecules as target nuclei. We use ecliptic coordinates (longitude $\lambda$ and latitude $\beta$).

We specify the WIMP velocities and the directions of nuclear recoil in a Cartesian coordinate system defined as follows:
as seen from the Earth, the $x$ axis points toward the position of the Sun at the vernal equinox, the $y$ axis toward the position of the Sun at the summer solstice, and the $z$ axis toward the North Pole of the ecliptic (which is the projection of the trajectory of the Sun onto the celestial sphere). The ecliptic longitude of the reference direction, $\lambda_{\rm axis}$, is the angular distance along the ecliptic from the vernal equinox to the base of the great circle containing ${\bf \hat n}$ and the pole of the ecliptic; it is measured eastwards in degrees from $0^{\circ}$ to $360^{\circ}$. The ecliptic latitude of the reference direction, $\beta_{\rm axis}$, is the angular distance north (from $0^{\circ}$ to $90^{\circ}$) or south (from $0^{\circ}$ to $-90^{\circ}$) of the ecliptic along the previously mentioned great circle; it is measured from the ecliptic to ${\bf \hat n}$. In terms of $\lambda_{\rm axis}$, $\beta_{\rm axis}$, we write
\begin{equation}
{\bf \hat n} = \left(\cos\beta_{\rm axis}\,\cos\lambda_{\rm axis},\,\,\cos\beta_{\rm axis}\,\sin\lambda_{\rm axis},\,\,\sin\beta_{\rm axis}\right)\,.
\end{equation}

In computing our rates we take a WIMP mass $m=60$ GeV, a WIMP-proton cross section $\sigma_{\rm p}=10^{-44}~{\rm cm}^2 = 10^{-8}~{\rm pb}$. We also neglect the nuclear form factor, i.e.\ take $\mathcal{F}_{n}(E)=1$. In this section, we further assume an ideal efficiency with zero threshold $\Eeff(E)=1$. In Sec.~\ref{Head-tail discrimination}, we present some results with a non-zero threshold for both spin-independent and spin-dependent cases.

\subsection{Streams of WIMPs}
\label{Stream of WIMPs}

We start by showing the results of applying the numerical method, Eq.~(\ref{dRdcostheRK}), to a simple model, namely a Maxwellian stream of WIMPs with an average velocity ${\bf V}$. For definiteness sake, we consider the case of a stream with ${\bf V} = (V,0,0)$, but our results apply to a generic ${\bf V}$.

\begin{figure}[t]
\centering
\includegraphics[width=\textwidth]{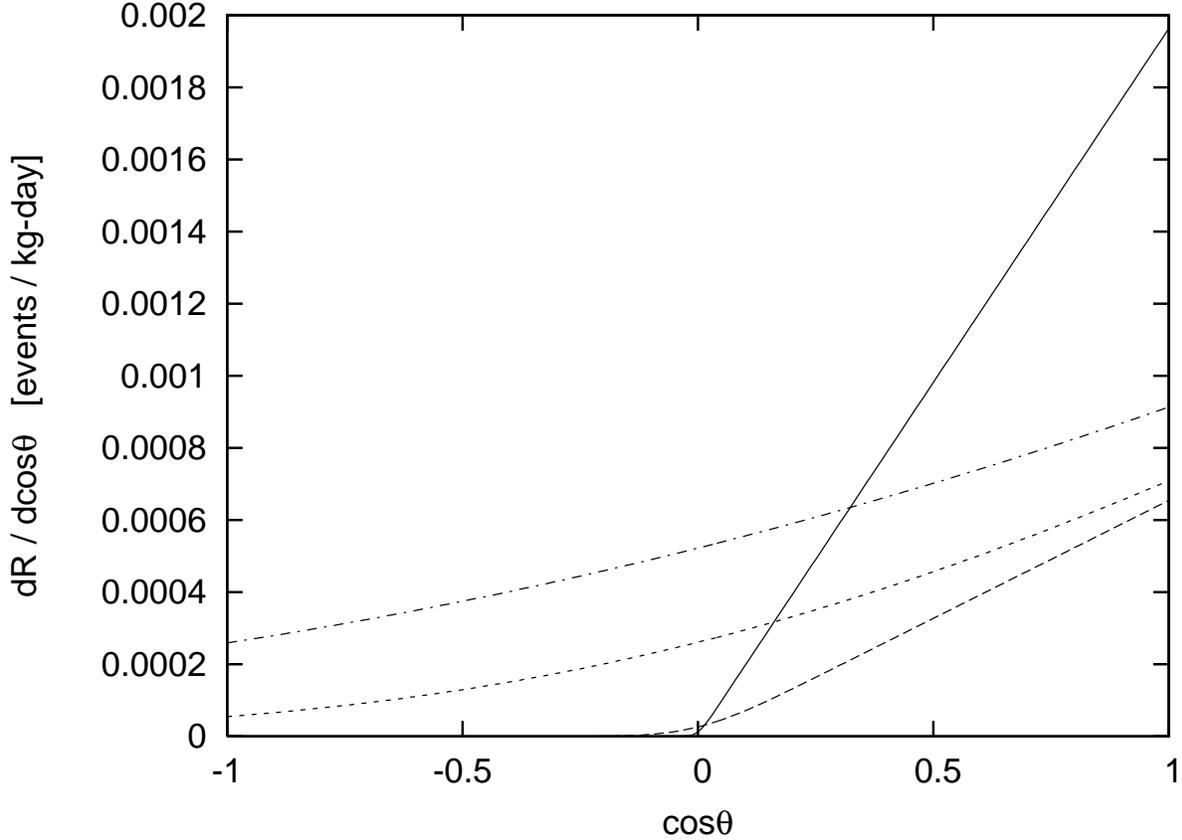}
\caption {The directional recoil rate ${dR}/{d\!\cos\theta}$ off a CS$_{2}$ target as a function of $\cos\theta$ for streams with average velocity ${\bf V}$ parallel to the reference direction ${\bf \hat n}$ for different $\sigma_{v}/V$ ratios; $\sigma_{v}/V={4.74}/{300}=0.0158$ (solid line), $\sigma_{v}/V={10}/{100}=0.1$ (dashed line), $\sigma_{v}/V={100}/{100}=1$ (dotted line), and $\sigma_{v}/V={200}/{100}=2$ (dashed-dotted line).}
\label{Fmvbx}
\end{figure}

First we fix the reference direction ${\bf \hat n}$ in the direction of ${\bf V}$, i.e.\ we take $(\lambda_{\rm axis},\beta_{\rm axis})=(0^{\circ},0^{\circ})$, and vary the ratio $\sigma_{v}/V$ of the velocity dispersion $\sigma_{v}$ to the magnitude of the average WIMP velocity $V$. Fig.~\ref{Fmvbx} shows the resulting  ${dR}/{d\!\cos\theta}$ as a function of $\cos\theta$ for streams with ${\sigma_{v}}/{V}={4.74}/{300}=0.0158$ (solid line), ${\sigma_{v}}/{V}={10}/{100}=0.1$ (dashed line), ${\sigma_{v}}/{V}={100}/{100}=1$ (dotted line), and ${\sigma_{v}}/{V}={200}/{100}=2$ (dashed-dotted line).

The solid line in Fig.~\ref{Fmvbx} shows that ${dR}/{d\!\cos\theta}$ peaks at $\theta = 0$ and almost vanishes for $\cos\theta\,\leq\,0$.  This behavior can be understood by considering a stream with zero velocity dispersion. All WIMPs in the stream move at the same velocity ${\bf V}$, so the velocity distribution function is
\begin{equation}
f({\bf v}) = \delta({\bf v} - {\bf V})\,.
\end{equation}
Its Radon transform follows from Eq.(\ref{eq:Rint}) as
\begin{equation}
{\hat f}(w, {\bf\hat w}) = \int {\delta ({\bf v} \cdot {\bf\hat w} - w) \,f({\bf v}) \,d^{3} \,v} = \delta ({\bf V} \cdot {\bf\hat w} - w),
\label{eq:fhatstream}
\end{equation}
With ${\bf \hat V} = V {\bf \hat n}$ and ${\bf\hat w} \cdot {\bf\hat n} = \cos\theta$, 
\begin{equation}
{\hat f}(w, {\bf\hat w})  = \delta(w - V\cos\theta).
\end{equation}
Inserting this relation into Eq.~(\ref{eq:In}), and using Eq.~(\ref{dE}), gives
\begin{eqnarray}
I_n & = &  \int \int {\hat f}(w_n,{\bf\hat w})\,dE\,d\phi \\
& = & 2 \pi \int_0^\infty \delta(w_n-V\cos\theta) \, dE \\
& = & \frac{4 \mu_n^2}{M_n c^2} \, 2 \pi \int_0^\infty \delta (w_n - V \cos\theta) w_n \, dw_n \\
& = & \frac{8 \pi \mu_n^2}{M_n c^2} \, V \cos\theta.
\end{eqnarray}
Therefore, for zero velocity dispersion, we have
\begin{equation}
\frac{dR}{d\cos\theta} = \left\{
\begin{array}{l l}
\displaystyle
 \left( \sum_n \frac{2\rho\sigma_{\rm p} \mu_n^2 C_n A_n^2}{\mu_{\rm p}^2 m M_n c^2}  \,V \right) \, \cos\theta\, , & \quad \mbox{if $\cos\theta$ $>$ 0\,,}\\
  0, & \quad \mbox{if $\cos\theta$ $\leq$ 0\,.}\\ \end{array} \right.
\end{equation}
For positive $\cos\theta$, ${dR}/{d\!\cos\theta}$ is a linearly increasing function of $\cos\theta$. Its maximum occurs in the forward direction at $\theta=0^{\circ}$. Away from the forward direction, the number of recoils decreases, and ${dR}/{d\!\cos\theta}$ drops. At negative $\cos\theta$ there are no recoils for $\sigma_{v}=0$, since momentum conservation forces all recoils to be in the forward direction. 

As the ratio $\sigma_{v}/V$ increases we can observe first a few  and then many recoils at $\cos\theta\,\leq\,0$, because of the effect of the relatively higher velocity dispersion $\sigma_{v}$ of the streams. This is illustrated by the dashed, dotted, and dashed-dotted lines in Fig.~\ref{Fmvbx}, some of which extend to $\cos\theta=-1$.

\begin{figure}[t]
\centering
\includegraphics[width=\textwidth]{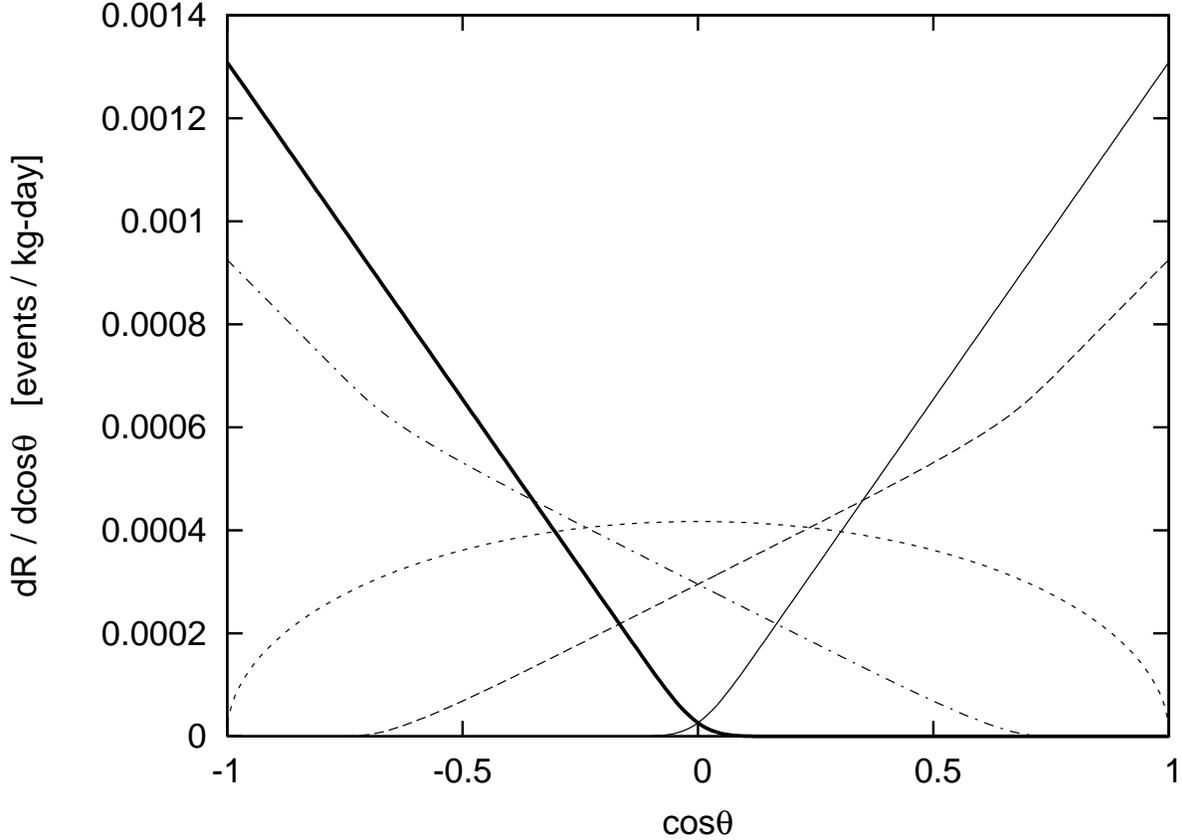}
\caption {The directional recoil rate  ${dR}/{d\!\cos\theta}$ off a CS$_{2}$ target for streams with $V = 200$ km/s, $\sigma_{v} = 10$km/s (thus ${\sigma_{v}}/{V}=0.05$) and various reference directions making an angle $\psi$ with ${\bf V}$ equal to: $0^{\circ}$ (solid line), $45^{\circ}$(dashed line), $90^{\circ}$ (dotted line), $135^{\circ}$ (dashed-dotted line), and $180^{\circ}$ (thick solid line).}
\label{FmvbxLamBet}
\end{figure}

Now we fix the velocity of the stream  and its dispersion, and vary the reference direction ${\bf \hat n}$. Since there are no other directions in this case, the rate $dR/d\!\cos\theta$ depends only on the angle $\psi$ between ${\bf V}$ and ${\bf \hat n}$, besides the ratio $\sigma_{v}/V$. For definiteness, we take ${\bf V} = (200,0,0)$ km/s and $\sigma_{v} = 10$ km/s (so the ratio ${\sigma_{v}}/{V}=0.05$). We start with the stream's average velocity ${\bf V}$ parallel to the reference direction $(\lambda_{\rm axis},\beta_{\rm axis})=(0^{\circ},0^{\circ})$. We keep the ecliptic latitude of the reference direction constant at $\beta_{\rm axis}=0^{\circ}$. We increase its ecliptic longitude  $\lambda_{\rm axis}$ in steps of $45^{\circ}$ until ${\bf V}$ is anti-parallel to ${\bf \hat n}$. Fig.~\ref{FmvbxLamBet} shows the corresponding ${dR}/{d\!\cos\theta}$ for $\psi=0^\circ, 45^\circ, 90^\circ, 135^\circ$, and $180^\circ$. We note the following.

First, the directional recoil rate ${dR}/{d\!\cos\theta}$ of streams parallel to the reference direction peaks in the direction opposite to the incoming WIMPs. For example, the solid line that peaks on the right is for WIMPs coming from the left and the thick solid line that peaks on the left is for WIMPs coming from the right. 

Second, for a given stream, if we take two reference directions that are opposite to each other, they form angles $\psi$ and $\pi-\psi$, respectively, with the stream's velocity ${\bf V}$. Since $\cos\psi=-\cos(\pi-\psi)$, their respective $dR/d\!\cos\theta$ distributions transform into each other under the substitution $\cos\theta\to-\cos\theta$. For example, the case $\psi=0^{\circ}$ (solid line in Fig.~\ref{FmvbxLamBet}) is the reflection about $\cos\theta=0$ of the case $\psi=180^{\circ}$ (thick solid line in Fig.~\ref{FmvbxLamBet}). Similarly, the case $\psi=45^{\circ}$ (dashed line in Fig.~\ref{FmvbxLamBet}) is the reflection of the case $\psi=135^{\circ}$ (dashed-dotted line in Fig.~\ref{FmvbxLamBet}).
For the same token, the case $\psi=90^\circ$ is the reflection of itself, i.e.\ it is symmetric under $\cos\theta\to-\cos\theta$.

\subsection{Standard dark halo}
\label{Standard dark halo}

Here we consider a flow of WIMPs according to the standard dark halo. In this model, the WIMPs are on average at rest relative to the Galaxy, and their velocity distribution is Maxwellian with velocity dispersion given by
\begin{equation}
\sigma_{v\,{\rm std}} = \frac{220\,{\rm km/s}}{\sqrt{2}}\,.
\end{equation}
The Local Standard of Rest (LSR) moves at $v_{\rm LSR} = 220$ km/s relative to the Galactic rest frame in the direction of the Galactic rotation, i.e.\ $(l_{\rm Gal.rot.},b_{\rm Gal.rot.}) = (90^{\circ},0^{\circ})$ in Galactic coordinates and $(\lambda_{\rm Gal.rot.},\beta_{\rm Gal.rot.}) = ({347\fdg340},{59\fdg574})$ in ecliptic coordinates.

Using the ecliptic Cartesian coordinate axes defined in the beginning of the current section, the average WIMP velocity components with respect to the Sun are
\begin{equation}
{\bf V}_{\rm std} = (-104.525,\,34.947,\,-196.836)\,{\rm km/s}\,.
\label{stdV}
\end{equation}
The average velocity ${\bf V}_{\rm std}$ points in the direction $(\lambda_{\rm std},\beta_{\rm std})=(161\fdg513,-60\fdg755)$. These values (${\bf V}_{\rm std}$ and its direction) are from Alenazi and Gondolo \cite{AG}.

In this subsection, we choose the reference direction ${\bf \hat n}$ along ${\bf V}_{\rm std}$, i.e.\ $(\lambda_{\rm axis},\beta_{\rm axis})=(161\fdg513,-60\fdg755)$.

\begin{figure}[t]
\centering
\includegraphics[width=\textwidth]{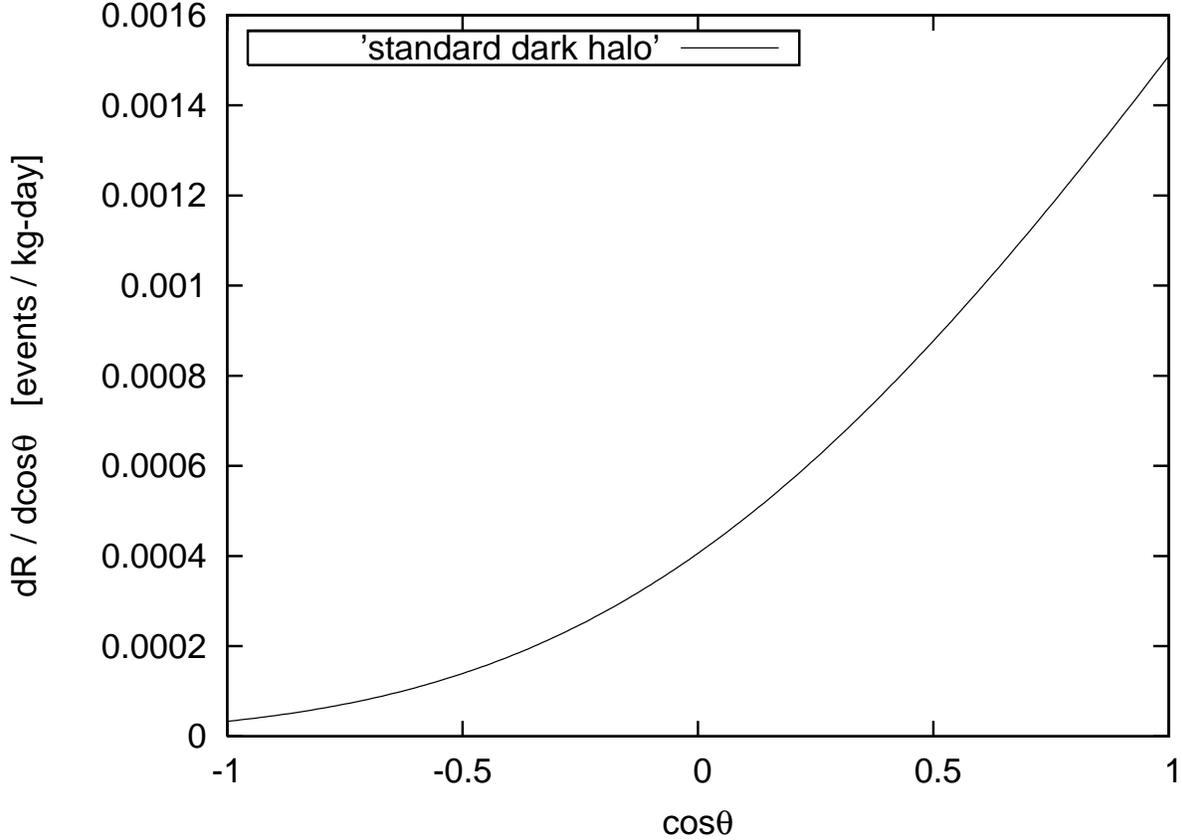}
\caption {The directional recoil rate  ${dR}/{d\!\cos\theta}$ off a CS$_{2}$ target for the standard dark halo for a reference direction ${\bf \hat n}$ in the direction $(\lambda_{\rm axis},\beta_{\rm axis})=(\lambda_{\rm std},\beta_{\rm std})=(161\fdg513,-60\fdg755)$.}
\label{FmStd}
\end{figure}

Applying the numerical method, Eq.~(\ref{dRdcostheRK}),  we see in Fig.~\ref{FmStd} that  ${dR}/{d\!\cos\theta}$  is an increasing function of $\cos\theta$. It peaks in the forward direction ($\theta=0^{\circ}$) because most of the recoils occur at $\theta=0^{\circ}$. Away from the forward direction, ${dR}/{d\!\cos\theta}$ decreases because fewer recoils result due to momentum conservation. The effect is similar to the dotted line in Fig.~\ref{Fmvbx}, whose ratio ${\sigma_{v}}/{V}=1$ is close to the ratio ${\sigma_{v\,{\rm std}}}/{V_{\rm std}}=\frac{155.567}{225.59}=0.689$ of the standard dark halo.  We further discuss the case of the standard dark halo (Fig.~\ref{FmStd}) in the following subsections.

\subsection{Sikivie's late-infall halo model (SLI streams)}
\label{Results for Sikivie's late-infall halo model}

Here we consider Sikivie's late-infall (SLI) halo model \cite{STW1,STW2,Si}. The SLI model is a self-similar axially symmetric infall model with net angular momentum and parameters adjusted to describe our Galaxy. In this model, many flows of collisionless DM particles are oscillating into and out of the Galaxy in pairs.  The first pair corresponds to particles passing through the Solar System from opposite sides of the Galactic Plane  for the first time, the second pair corresponds to particles passing for the second time, and so on (from the fifth pair onward, the flows in a pair come on the Galactic Plane but one inwards and the other outwards). Table \ref{table:pairs} lists the densities $\rho_{i}$ and velocities ${\bf V}_{i}$ of the first 20 pairs of SLI streams in our ecliptic coordinate system. These values are taken from Ref.~\cite{Sikivie}, where they were given in the Galactic coordinate system (see also \cite{GG}).

\begin{table}[t]
\caption{Densities and velocities of the first 20 pairs ($\pm$) of SLI streams in the ecliptic coordinate system. The values are from Ref.~\cite{Sikivie} where they were given in the Galactic coordinate system.}
\centering
\begin{tabular}{c c rrr}
\hline\hline \\[-5ex] 
$i$ \,\,\,\,\,\,\,\,\,\,\,\,\,\,& $\rho_{i}$ \, \, \, \, \, & $V_{ix}$ \, \, \, \, \, & $V_{iy}$ \, \, \, \, \, & $V_{iz}$ \, \, \, \, \, \\ [-1.5ex]
 \, \, \, \, \,  & ($10^{-26}\,{\rm g/cm^{3}}$) \, \, \, \, \, & (km/s) \, \, \, & (km/s) \, \, \, & (km/s) \, \, \, \\ [0ex]
\hline \\[-5ex] 
& & -560 \, \, \, & 19 \, \, \, & 225 \, \, \, \\[-2.6ex]
\raisebox{1.5ex}{$1^\pm$ \, \, \, \, \, } & \raisebox{1.5ex}{0.4 \, \, \, \, \, } & 490 \, \, \,  & 20 \, \, \, & -377 \, \, \, \\[-1.8ex]

& & -417 \, \, \, & 6.5 \, \, \, & 274 \, \, \, \\[-2.6ex]
\raisebox{1.5ex}{$2^\pm$ \, \, \, \, \, } & \raisebox{1.5ex}{1.0 \, \, \, \, \, } & 460 \, \, \,  & 6.8 \, \, \, & -228 \, \, \, \\[-1.8ex]

& & -270 \, \, \, & -4.0 \, \, \, & 299 \, \, \, \\[-2.6ex]
\raisebox{1.5ex}{$3^\pm$ \, \, \, \, \, } & \raisebox{1.5ex}{2.0 \, \, \, \, \, } & 407 \, \, \,  & -3.8 \, \, \, & -89 \, \, \, \\[-1.8ex]

& & -95 \, \, \, & -14 \, \, \, & 302 \, \, \, \\[-2.6ex]
\raisebox{1.5ex}{$4^\pm$ \, \, \, \, \, } & \raisebox{1.5ex}{6.3 \, \, \, \, \, } & 321 \, \, \,  & -14 \, \, \, & 63 \, \, \, \\[-1.8ex]

& & 102 \, \, \, & -203 \, \, \, & 164 \, \, \, \\[-2.6ex]
\raisebox{1.5ex}{$5^\pm$ \, \, \, \, \, } & \raisebox{1.5ex}{9.2 \, \, \, \, \, } & 123 \, \, \,  & 175 \, \, \, & 201 \, \, \, \\[-1.8ex]

& & 55 \, \, \, & -298 \, \, \, & 81 \, \, \, \\[-2.6ex]
\raisebox{1.5ex}{$6^\pm$ \, \, \, \, \, } & \raisebox{1.5ex}{2.9 \, \, \, \, \, } & 87 \, \, \,  & 289 \, \, \, & 138 \, \, \, \\[-1.8ex]

& & 21 \, \, \, & -325 \, \, \, & 21 \, \, \, \\[-2.6ex]
\raisebox{1.5ex}{$7^\pm$ \, \, \, \, \, } & \raisebox{1.5ex}{1.9 \, \, \, \, \, } & 57 \, \, \,  & 331 \, \, \, & 85 \, \, \, \\[-1.8ex]

& & -0.2 \, \, \, & -341 \, \, \, & -15 \, \, \, \\[-2.6ex]
\raisebox{1.5ex}{$8^\pm$ \, \, \, \, \, } & \raisebox{1.5ex}{1.4 \, \, \, \, \, } & 38 \, \, \,  & 355 \, \, \, & 52 \, \, \, \\[-1.8ex]

& & -18 \, \, \, & -342 \, \, \, & -46 \, \, \, \\[-2.6ex]
\raisebox{1.5ex}{$9^\pm$ \, \, \, \, \, } & \raisebox{1.5ex}{1.1 \, \, \, \, \, } & 21 \, \, \,  & 364 \, \, \, & 23 \, \, \, \\[-1.8ex]

& & -30 \, \, \, & -339 \, \, \, & -67 \, \, \, \\[-2.6ex]
\raisebox{1.5ex}{$10^\pm$ \, \, \, \, \, } & \raisebox{1.5ex}{1.0 \, \, \, \, \, } & 8.8 \, \, \,  & 367 \, \, \, & 1.3 \, \, \, \\[-1.8ex]

& & -40 \, \, \, & -337 \, \, \, & -84 \, \, \, \\[-2.6ex]
\raisebox{1.5ex}{$11^\pm$ \, \, \, \, \, } & \raisebox{1.5ex}{0.9 \, \, \, \, \, } & -1.0 \, \, \,  & 369 \, \, \, & -16 \, \, \, \\[-1.8ex]

& & -50 \, \, \, & -330 \, \, \, & -101 \, \, \, \\[-2.6ex]
\raisebox{1.5ex}{$12^\pm$ \, \, \, \, \, } & \raisebox{1.5ex}{0.8 \, \, \, \, \, } & -11 \, \, \,  & 366 \, \, \, & -34 \, \, \, \\[-1.8ex]

& & -57 \, \, \, & -323 \, \, \, & -114 \, \, \, \\[-2.6ex]
\raisebox{1.5ex}{$13^\pm$ \, \, \, \, \, } & \raisebox{1.5ex}{0.7 \, \, \, \, \, } & -19 \, \, \,  & 363 \, \, \, & -47 \, \, \, \\[-1.8ex]

& & -64 \, \, \, & -316 \, \, \, & -126 \, \, \, \\[-2.6ex]
\raisebox{1.5ex}{$14^\pm$ \, \, \, \, \, } & \raisebox{1.5ex}{0.6 \, \, \, \, \, } & -27 \, \, \,  & 359 \, \, \, & -61 \, \, \, \\[-1.8ex]

& & -68 \, \, \, & -305 \, \, \, & -134 \, \, \, \\[-2.6ex]
\raisebox{1.5ex}{$15^\pm$ \, \, \, \, \, } & \raisebox{1.5ex}{0.6 \, \, \, \, \, } & -32 \, \, \,  & 351 \, \, \, & -70 \, \, \, \\[-1.8ex]

& & -73 \, \, \, & -299 \, \, \, & -142 \, \, \, \\[-2.6ex]
\raisebox{1.5ex}{$16^\pm$ \, \, \, \, \, } & \raisebox{1.5ex}{0.55 \, \, \, \, \, } & -37 \, \, \,  & 347 \, \, \, & -79 \, \, \, \\[-1.8ex]

& & -78 \, \, \, & -293 \, \, \, & -150 \, \, \, \\[-2.6ex]
\raisebox{1.5ex}{$17^\pm$ \, \, \, \, \, } & \raisebox{1.5ex}{0.5 \, \, \, \, \, } & -42 \, \, \,  & 343 \, \, \, & -88 \, \, \, \\[-1.8ex]

& & -80 \, \, \, & -283 \, \, \, & -153 \, \, \, \\[-2.6ex]
\raisebox{1.5ex}{$18^\pm$ \, \, \, \, \, } & \raisebox{1.5ex}{0.5 \, \, \, \, \, } & -45 \, \, \,  & 334 \, \, \, & -94 \, \, \, \\[-1.8ex]

& & -82 \, \, \, & -277 \, \, \, & -157 \, \, \, \\[-2.6ex]
\raisebox{1.5ex}{$19^\pm$ \, \, \, \, \, } & \raisebox{1.5ex}{0.45 \, \, \, \, \, } & -48 \, \, \,  & 329 \, \, \, & -98 \, \, \, \\[-1.8ex]

& & -84 \, \, \, & -271 \, \, \, & -161 \, \, \, \\[-2.6ex]
\raisebox{1.5ex}{$20^\pm$ \, \, \, \, \, } & \raisebox{1.5ex}{0.45 \, \, \, \, \, } & -51 \, \, \,  & 325 \, \, \, & -103 \, \, \, \\[0ex]

\hline\hline
\end{tabular}
\label{table:pairs}
\end{table}

The conversion from the Galactic frame to ecliptic frame proceeds as follows. In the Galactic coordinate system, $X$ points toward the Galactic Center, $Y$ toward the direction of Galactic rotation, and $Z$ toward the North Galactic Pole. Our ecliptic coordinate system assumes $x$ pointing toward the vernal equinox, $y$ toward the summer solstice, and $z$ toward the North Pole of the ecliptic. 
We take the Galactic components of the solar motion to be \cite{WJ} $U = 10.00 \pm 0.36$ km/s (radially inwards), $V = 5.25 \pm 0.62$ km/s (in the direction of Galactic rotation) and $W = 7.17 \pm 0.38$ km/s (vertically upwards). The velocities ${\bf V}_{i}^{S}$ of the SLI streams relative to the Sun follow, in Galactic coordinate system, as
\begin{equation}
V_{iX}^{S} = V_{iX}^{G}-U\,,
\end{equation}
\begin{equation}
V_{iY}^{S} = V_{iY}^{G}-V-v_{\rm LSR}\,,
\end{equation}
\begin{equation}
V_{iZ}^{S} = V_{iZ}^{G}-W\,,
\end{equation}
where ${\bf V}_{i}^{G}$ are the velocities relative to the Galaxy (and $v_{\rm LSR}=220$ km/s, as in the previous subsection). Then we convert $V_{iX}^{S}$,\,$V_{iY}^{S}$,\,and $V_{iZ}^{S}$ from the Galactic coordinate system to our ecliptic coordinate system, and obtain the velocities of SLI streams $V_{ix}$,\,$V_{iy}$,\,and $V_{iz}$ listed in Table \ref{table:pairs}.

In the SLI model, the WIMP velocity distribution function is given by
\begin{equation}
f_{\rm SLI}({\bf v}) = \frac{1}{\rho}\,\sum_{i}\,\rho_{i}\,\delta({\bf v}-{\bf V}_{i})\,.
\label{fSLI}
\end{equation}
Following \cite{GG}, we assume a Gaussian distribution of velocities for each flow with velocity dispersion
$\sigma_{v\,{\rm SLI}} = \frac{v_{0}}{\sqrt{2}}$ with $v_{0} = 30\,{\rm km/s}$.
We have 
\begin{equation}
f_{\rm SLI}({\bf v}) = \frac{1}{\rho}\,\sum_{i}\,\frac{\rho_{i}}{(2 \pi \sigma_{v\,{\rm SLI}}^2)^{3/2}} \exp\!{\left(- {\frac{\left| {\bf v} - {\bf V}_{i} \right|^2}{2 \sigma_{v\,{\rm SLI}}^2}}\right)}\,.
\label{fSLI1}
\end{equation}
Now, the recoil momentum distribution is
\begin{equation}
{\hat f}_{\rm SLI}(w, {\bf\hat w}) = \frac{1}{\rho}\,\sum_{i}\,\frac{\rho_{i}}{(2 \pi \sigma_{v\,{\rm SLI}}^2)^{1/2}} \exp\!{\left(- {\frac{[w-{\bf\hat w} \cdot {\bf V}_{i}]^{2}}{2 \sigma_{v\,{\rm SLI}}^2}}\right)}\,.
\label{fhSLI}
\end{equation}

\begin{figure}[t]
\centering
\includegraphics[width=\textwidth]{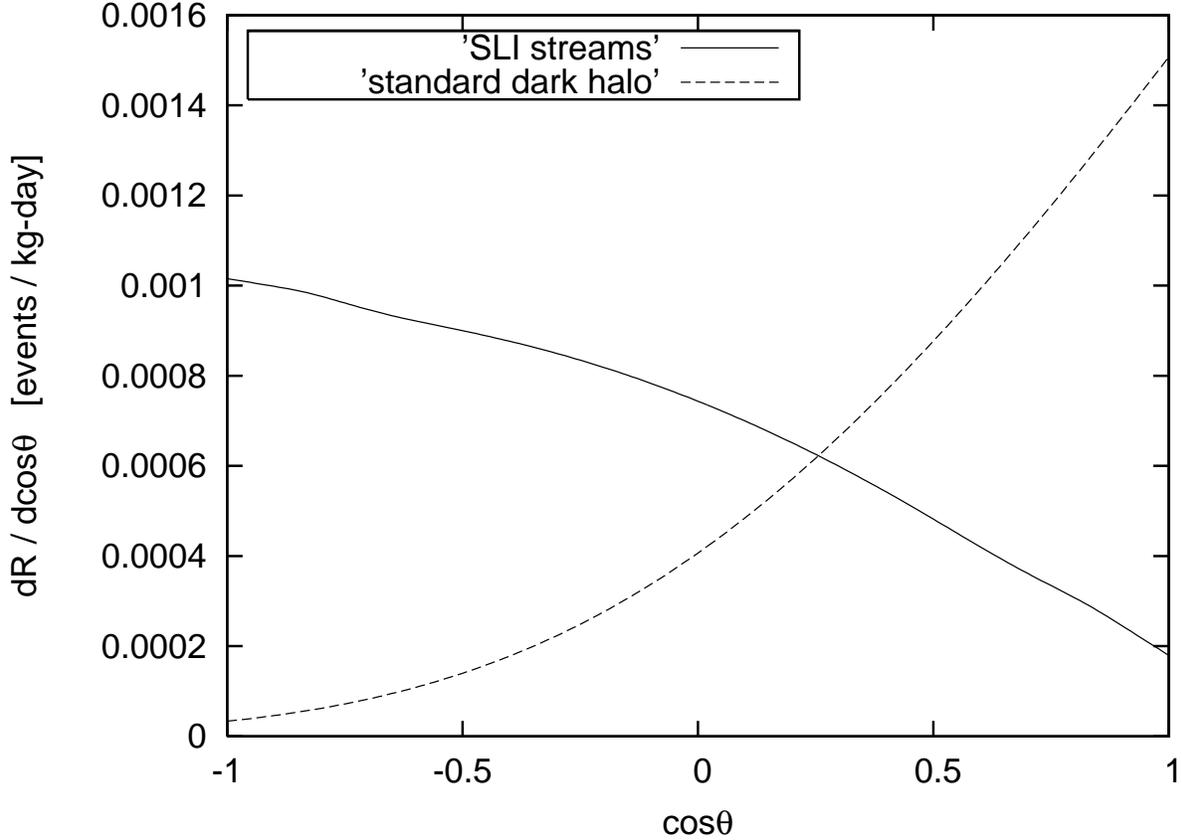}
\caption {The directional recoil rate ${dR}/{d\!\cos\theta}$ off a CS$_{2}$ target for SLI streams (solid line) and the standard dark halo (dashed line) for a reference direction opposite to the direction of Galactic rotation, $(\lambda_{\rm axis},\beta_{\rm axis})=(167\fdg340,-59\fdg574)$.}
\label{dRdcosRKSLIstd}
\end{figure}

We fix the reference direction opposite to the direction of Galactic rotation, $(l_{\rm axis},b_{\rm axis})=(270^{\circ},0^{\circ})$ in Galactic coordinates or $(\lambda_{\rm axis},\beta_{\rm axis})=(167\fdg340,-59\fdg574)$ in ecliptic coordinates. We used the numerical method, Eq.~(\ref{dRdcostheRK}), to calculate the directional recoil rate ${dR}/{d\!\cos\theta}$ for SLI streams. The result is shown in Fig.~\ref{dRdcosRKSLIstd} by the solid line. For comparison, we also show the result from the standard dark halo (dashed line), recomputed for the new reference direction.

We see in Fig.~\ref{dRdcosRKSLIstd} that ${dR}/{d\cos\theta}|_{\rm SLI}$ (solid line) peaks in the direction opposite to the case of ${dR}/{d\cos\theta}|_{\rm std}$ (dashed line). This is because the average velocity of the SLI streams points in a direction roughly opposite to that of standard dark halo's. Indeed, using Table \ref{table:pairs}, the average (weighted) velocity ${\bf \bar V}_{\rm SLI}$ of the SLI streams is
\begin{equation}
{\bf \bar V}_{\rm SLI} = (57.534,\,-1.465,\,85.994)\,{\rm km/s}\,.
\label{SLIV}
\end{equation}
Comparison with Eq.~(\ref{stdV}) shows that ${\bf V}_{\rm std}$ and ${\bf \bar V}_{\rm SLI}$ form an angle of 170$^\circ$. 

\subsection{Anisotropic logarithmic-ellipsoidal models}
\label{Results for anisotropic velocity dispersion}

Many observations and numerical simulations show that galaxy halos are better approximated by triaxial models with anisotropic velocity distributions (e.g. Moore {\it et al.} \cite{Mooreetal}, Helmi, White, and Springel \cite{HWS}, and Green \cite{Green}). Green \cite{Green} examined the effect of triaxial and anisotropic halo models on the exclusion limits from WIMP direct detection searches and found that such models lead to non-negligible changes in the exclusion limits. Helmi, White, and Springel \cite{HWS} found that the expected signal for the fastest moving DM particles in direct detection experiments is highly anisotropic. In this subsection we investigate the nuclei's directional recoil rate ${dR}/{d\!\cos\theta}$ in specific cases of anisotropic logarithmic-ellipsoidal models.

Evans, Carollo, and de Zeeuw \cite{ECdeZ} provided analytic solutions to the Jeans equations for the logarithmic-ellipsoidal model, which is the simplest triaxial generalization of the isothermal sphere, under the assumption of conical alignment of the velocity ellipsoid. Their anisotropic velocity distribution at the Sun's position can be approximated by an anisotropic Gaussian, which in the Galactic rest frame where the WIMPs are on average at rest is given  by
\begin{equation}
f({\bf v}) = \frac{1}{(2\,\pi)^{3/2}\,\sigma_{vX}\,\sigma_{vY}\,\sigma_{vZ}}\exp\!{\left(- \frac{v_{X}^{2}}{2\,\sigma_{vX}^{2}} - \frac{v_{Y}^{2}}{2\,\sigma_{vY}^{2}} - \frac{v_{Z}^{2}}{2\,\sigma_{vZ}^{2}}\right)}\,.
\label{anisof}
\end{equation}
Here $X$ points toward the Galactic Center, $Y$ toward the direction of Galactic rotation, and $Z$ toward the North Galactic Pole. In this frame, the velocity dispersion matrix $\bm\sigma_{v}^{2}$ is diagonal.

In one of the Evans, Carollo, and de Zeeuw models, the Solar System is on the long (major) axis of the halo density ellipsoid, and
\begin{eqnarray}
\sigma_{vX}^{2} & = & \frac{v_{0}^{2}}{(2+\gamma)(p^{-2}+q^{-2}-1)}\,,
\label{sigmax2}
\\
\sigma_{vY}^{2} & = & \frac{v_{0}^{2}(2\,p^{-2}-1)}{2(p^{-2}+q^{-2}-1)}\,,
\\
\sigma_{vZ}^{2} & = & \frac{v_{0}^{2}(2\,q^{-2}-1)}{2(p^{-2}+q^{-2}-1)}\,.
\end{eqnarray}
In another of their models, the Solar System is on the intermediate (minor) axis of the halo density ellipsoid, and
\begin{eqnarray}
\sigma_{vX}^{2} & = & \frac{v_{0}^{2}\,p^{-4}}{(2+\gamma)(1+q^{-2}-p^{-2})}\,,
\\
\sigma_{vY}^{2} & = & \frac{v_{0}^{2}(2-p^{-2})}{2(1+q^{-2}-p^{-2})}\,,
\\
\sigma_{vZ}^{2} & = & \frac{v_{0}^{2}(2\,q^{-2}-p^{-2})}{2(1+q^{-2}-p^{-2})}\,.
\label{sigmaz2}
\end{eqnarray}
Here $p$ and $q$ are constants used to describe the axis ratios of the density ellipsoid, and $\gamma$ is a constant used to describe the velocity anisotropy. In the spherical limit ($p = q = 1$, $\gamma = 0$), all the velocity dispersion components are equal, and the logarithmic-ellipsoidal model reduces to the isothermal sphere.


An anisotropic Gaussian velocity distribution can be written in matrix form as
\begin{equation}
f({\bf v}) = \frac{1}{(2\pi)^{3/2}\,(\det \bm\sigma_{v}^2)^{1/2}} \exp\left[ - \frac{1}{2} ({\bf v}-{\bf V})^T \bm\sigma_{v}^{-2} ({\bf v}-{\bf V}) \right].
\end{equation}
Its Radon transform has been found to be \cite{Gon}
\begin{equation}
\hat f(w, {\bf \hat w}) = \frac{1}{(2\,\pi\,{\hat{\bf w}}^{T}\,\bm\sigma_{v}^{2}\,{\hat{\bf w}})^{1/2}}\exp\!{\left(- \frac{[w - {\bf {\hat w}} \cdot {\bf V}]^{2}}{2\,{\hat{\bf w}}^{T}\,\bm\sigma_{v}^{2}\,{\hat{\bf w}}}\right)}\,.
\label{Gondolofhat}
\end{equation}

Since the WIMPs are assumed to be on average at rest in the Galactic frame, the same average WIMP velocity ${\bf V}$  as for the standard dark halo ${\bf V}_{\rm std}$ (see Eq.~(\ref{stdV})) applies to these anisotropic models. Since the velocity dispersion matrix $\bm\sigma_{v}^{2}$ is diagonal in the Galactic frame, the principal axes of the velocity dispersion ellipsoid are aligned with the axes $X$,\,$Y$,\, and $Z$ (see Eq.~(\ref{anisof})). In matrix form, the exponent of Eq.~(\ref{anisof}) can be written as
\begin{equation}
	- \frac{1}{2}
	\begin{pmatrix}
		{v}_{X} & {v}_{Y} & {v}_{Z} 
	\end{pmatrix} 
	\begin{pmatrix}
		\frac{1}{\sigma_{vX}^2} & 0 & 0 \\
		 0 & \frac{1}{\sigma_{vY}^2} & 0 \\
        	 0 & 0 & \frac{1}{\sigma_{vZ}^2}
        \end{pmatrix}
	\begin{pmatrix}
		{v}_{X} \\
		{v}_{Y} \\
                {v}_{Z}
	\end{pmatrix}\,,
\end{equation}
and the velocity dispersion tensor is
\begin{equation}
	{\bm\sigma_{v}^{2}} = \begin{pmatrix}
		\sigma_{vX}^{2} & 0 & 0 \\
		 0 & \sigma_{vY}^{2} & 0 \\
        	 0 & 0 & \sigma_{vZ}^{2}
	\end{pmatrix}\,.
\end{equation}
To calculate ${\hat{\bf w}}^{T}\,\bm\sigma_{v}^{2}\,{\hat{\bf w}}$ in Eq.~(\ref{Gondolofhat}), we write
\begin{equation}
	{\hat{\bf w}}^{T}\,\bm\sigma_{v}^{2}\,{\hat{\bf w}} = 
	\begin{pmatrix}
		{\hat w}_{X} & {\hat w}_{Y} & {\hat w}_{Z} \\
	\end{pmatrix} 
	\begin{pmatrix}
	        \sigma_{vX}^{2} & 0 & 0 \\
		 0 & \sigma_{vY}^{2} & 0 \\
        	 0 & 0 & \sigma_{vZ}^{2}
	\end{pmatrix} 
	\begin{pmatrix}
		{\hat w}_{X} \\
		{\hat w}_{Y} \\
                {\hat w}_{Z}
	\end{pmatrix} = 
	\sigma_{vX}^{2}\,{\hat w}_{X}^{2} + \sigma_{vY}^{2}\,{\hat w}_{Y}^{2} + \sigma_{vZ}^{2}\,{\hat w}_{Z}^{2}\,.
\end{equation}
Thus, the anisotropic recoil momentum distribution is given by
\begin{equation}
{\hat f}(w, {\bf {\hat w}}) = \frac{1}{\sqrt{2\,\pi(\sigma_{vX}^{2}\,{\hat w}_{X}^{2} + \sigma_{vY}^{2}\,{\hat w}_{Y}^{2} + \sigma_{vZ}^{2}\,{\hat w}_{Z}^{2})}}\exp\!{\left(- \frac{[w - {\bf {\hat w}} \cdot {\bf V}]^{2}}{2(\sigma_{vX}^{2}\,{\hat w}_{X}^{2} + \sigma_{vY}^{2}\,{\hat w}_{Y}^{2} + \sigma_{vZ}^{2}\,{\hat w}_{Z}^{2})}\right)}\,,
\label{anisof1}
\end{equation}
where only the components of $\sigma_{v}$ and $\hat w$ are now in Galactic coordinates.

\begin{figure}[t]
\centering
\includegraphics[width=\textwidth]{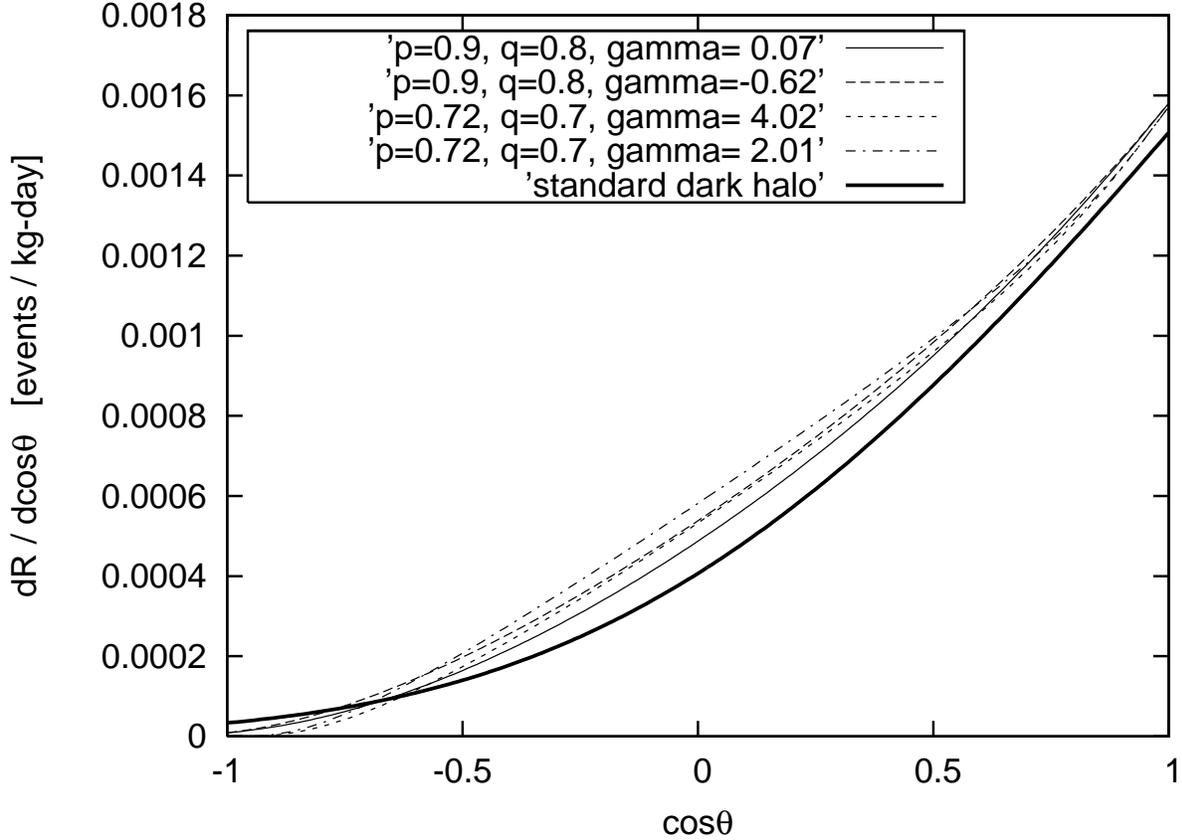}
\caption {The directional recoil rate  ${dR}/{d\!\cos\theta}$ off a CS$_{2}$ target as a function of $\cos\theta$ for logarithmic-ellipsoidal anisotropic models in which the Solar System is on the minor axis of the density ellipsoid (solid, dashed, dotted, and dashed-dotted lines, with parameters given in the legend). Also shown is the case of the isotropic standard dark halo (thick solid line). The reference direction ${\bf \hat n}$ is opposite to the direction of the Galactic rotation, $(\lambda_{\rm axis},\beta_{\rm axis})=(167\fdg340,-59\fdg574)$.}
\label{dRdcosRKanisoInter}
\end{figure}
\begin{figure}[t]
\centering
\includegraphics[width=\textwidth]{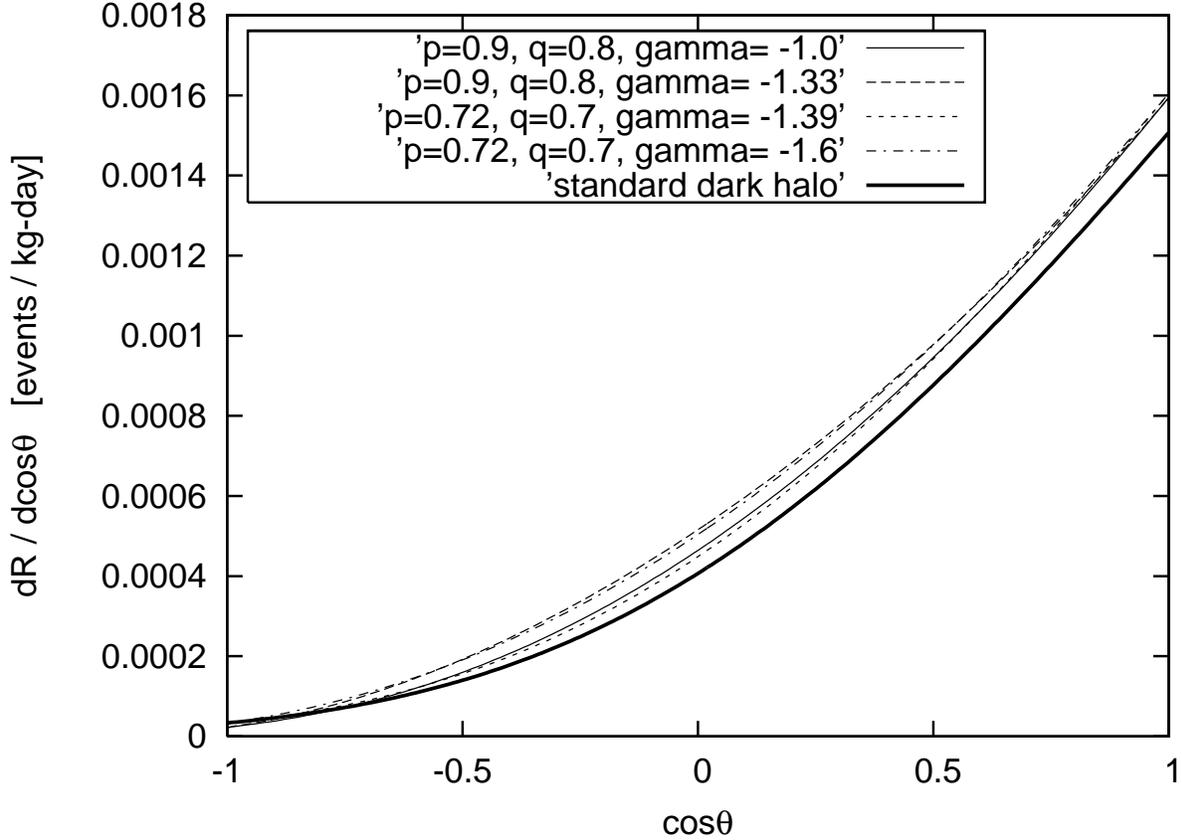}
\caption {Same as Fig.~\ref{dRdcosRKanisoInter} but for logarithmic-ellipsoidal models in which the Solar System is on the major axis of the density ellipsoid.}
\label{dRdcosRKanisoMaj}
\end{figure}

We fix the reference direction to be opposite to the direction of Galactic rotation, $(l_{\rm axis},b_{\rm axis})=(270^{\circ},0^{\circ})$ in Galactic coordinates or $(\lambda_{\rm axis},\beta_{\rm axis})=(167\fdg340,-59\fdg574)$ in the ecliptic coordinates. We consider eight anisotropic models (i.e.\ different values of $p$, $q$, and $\gamma$) taken from Green \cite{Green}. Using the numerical method, Eq.~(\ref{dRdcostheRK}), the results are shown in Fig.~\ref{dRdcosRKanisoInter} in the case when the Solar System is on the minor axis of the density ellipsoid and in Fig.~\ref{dRdcosRKanisoMaj} in the case when the Solar System is on the major axis of the density ellipsoid.  For comparison, in both figures we also show the case of the isotropic standard dark halo discussed in Sec.~\ref{Standard dark halo}  (thick solid line). The values of $p$, $q$, and $\gamma$ in Figs.~\ref{dRdcosRKanisoInter} and ~\ref{dRdcosRKanisoMaj} are taken from Table I in \cite{Green}.

In Figs.~\ref{dRdcosRKanisoInter} and ~\ref{dRdcosRKanisoMaj}, ${dR}/{d\!\cos\theta}$ is an increasing function of $\cos\theta$. It is maximum in the forward direction ($\theta=0^{\circ}$) because of the direction of the average WIMP velocity is similar to that of the standard halo's. We see from Figs.~\ref{dRdcosRKanisoInter} and ~\ref{dRdcosRKanisoMaj} that the behavior of the eight anisotropic models (solid, dashed, dotted, and dashed-dotted lines) resembles that of the isotropic standard dark halo (thick solid line). The differences between these models in Figs.~\ref{dRdcosRKanisoInter} and ~\ref{dRdcosRKanisoMaj} arise from the different velocity dispersions $\sigma_{v}$ of these models resulting from different values of the parameter $p$, $q$, and $\gamma$ in Eqs.~(\ref{sigmax2}--\ref{sigmaz2}). This also means different ratios $\sigma_{v}/V$ for each model.

\section{``Folded'' directional recoil rate}
\label{Head-tail discrimination}

In this section, we will discuss the consequences of a lack of head-tail discrimination in WIMP direct detectors. We use the ``folded'' directional recoil rate ${dR}/{d|\!\cos\theta|}$ defined in Eq.~(\ref{dRdcosabs}). We also discuss the exposures required to distinguish the expected WIMP signals from an isotropic noise, and present results for both an ideal zero-threshold detector and a detector with a finite energy threshold of 20 keV. We consider a CS$_2$ target (spin-independent and spin-dependent, which is zero) and a CF$_4$ target (assuming a spin-dependent cross section only).

\begin{figure}[t]
\centering
\includegraphics[width=\textwidth]{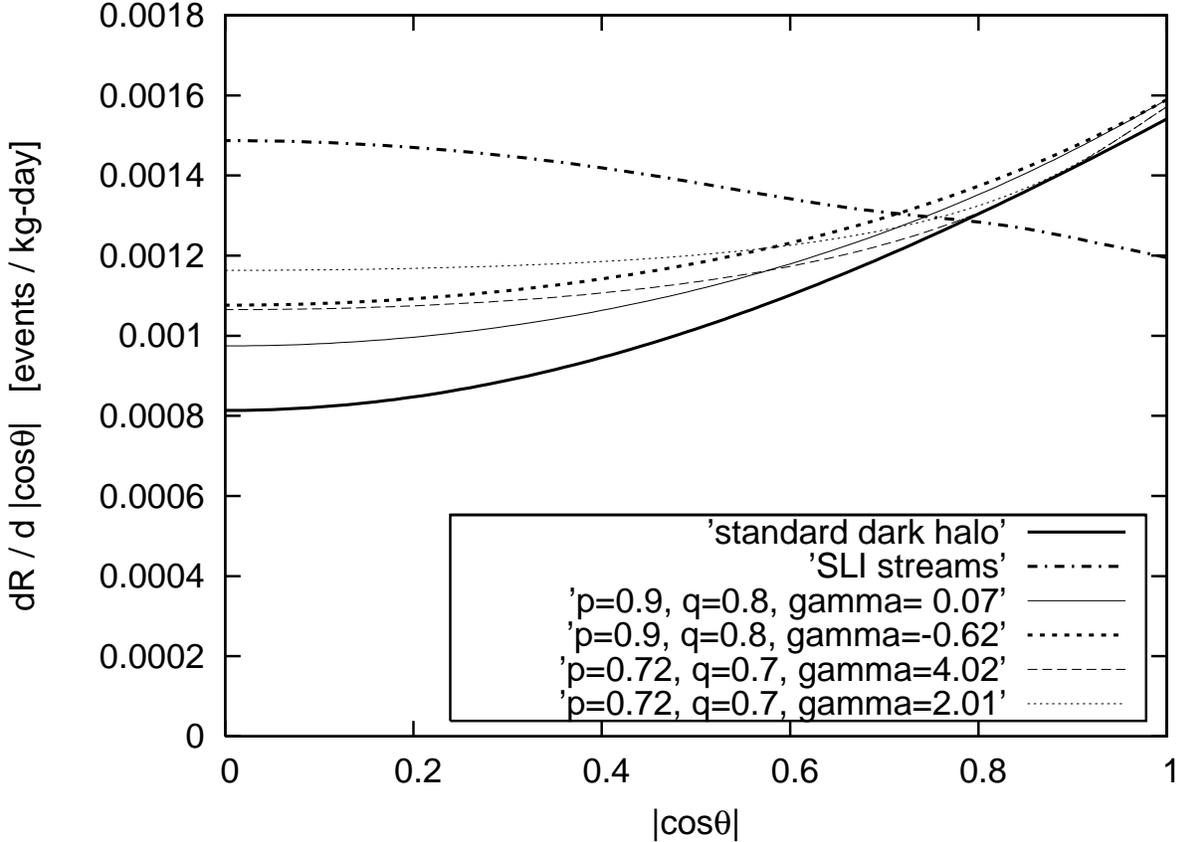}
\caption {The ``folded'' directional recoil rate  ${dR}/{d|\!\cos\theta|}$ off a CS$_{2}$ target for the standard dark halo (thick solid line), SLI streams (thick dashed-dotted line), and four logarithmic-ellipsoidal anisotropic models in which the Solar System is on the minor axis of the density ellipsoid (solid, thick dotted, dashed, and dotted lines, with parameters given in the legend). In all cases, the reference direction is opposite to the direction of the Galactic rotation, $(\lambda_{\rm axis},\beta_{\rm axis})=(167\fdg340,-59\fdg574)$.}
\label{absCosSLIstdAnisoInter}
\end{figure}

\begin{figure}[t]
\centering
\includegraphics[width=\textwidth]{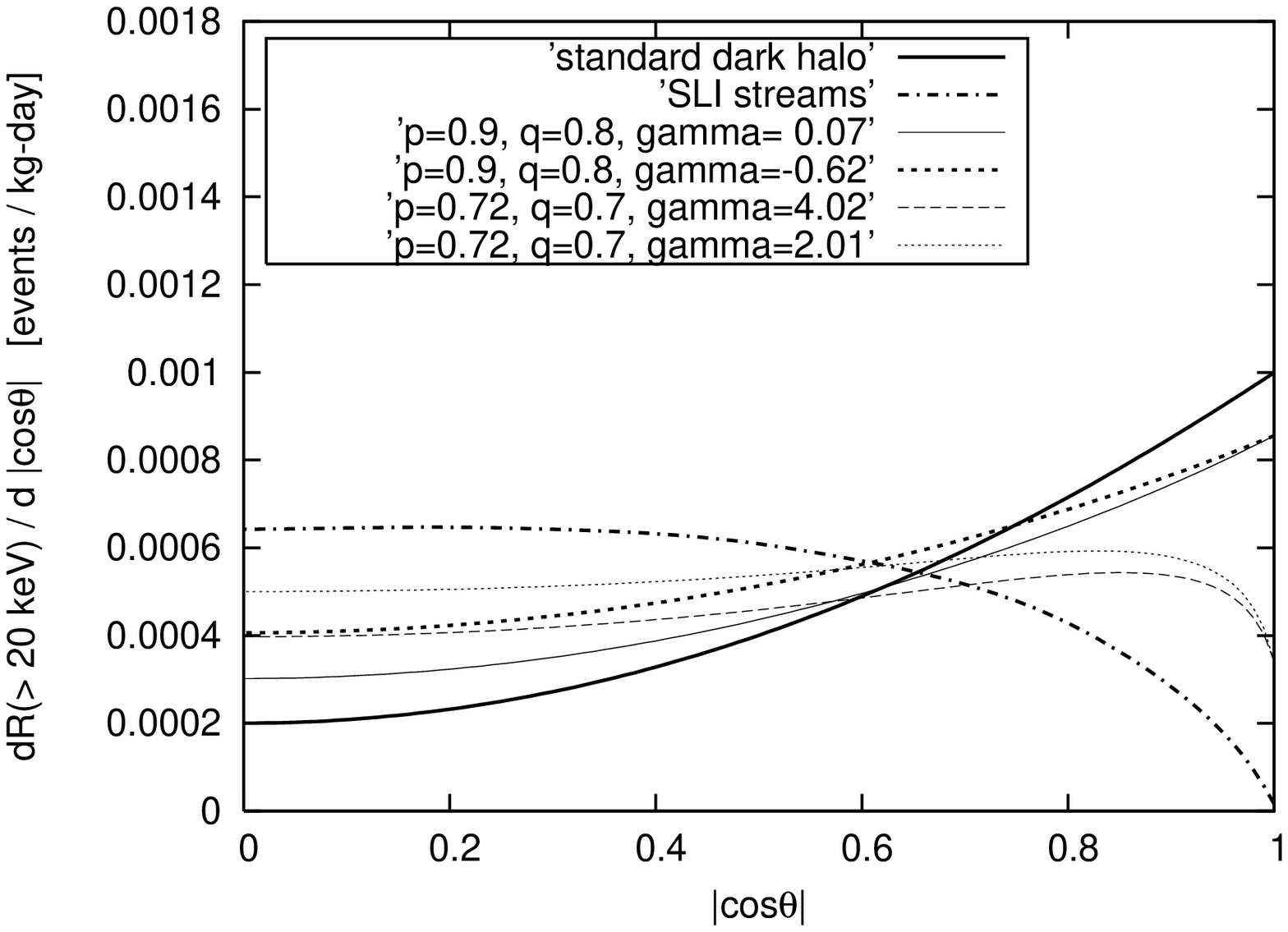}
\caption {Same as Fig.~\ref{absCosSLIstdAnisoInter} but using a detector with threshold energy equal to 20 keV.}
\label{absCosSLIstdAnisoInterTh}
\end{figure}

\begin{figure}[t]
\centering
\includegraphics[width=\textwidth]{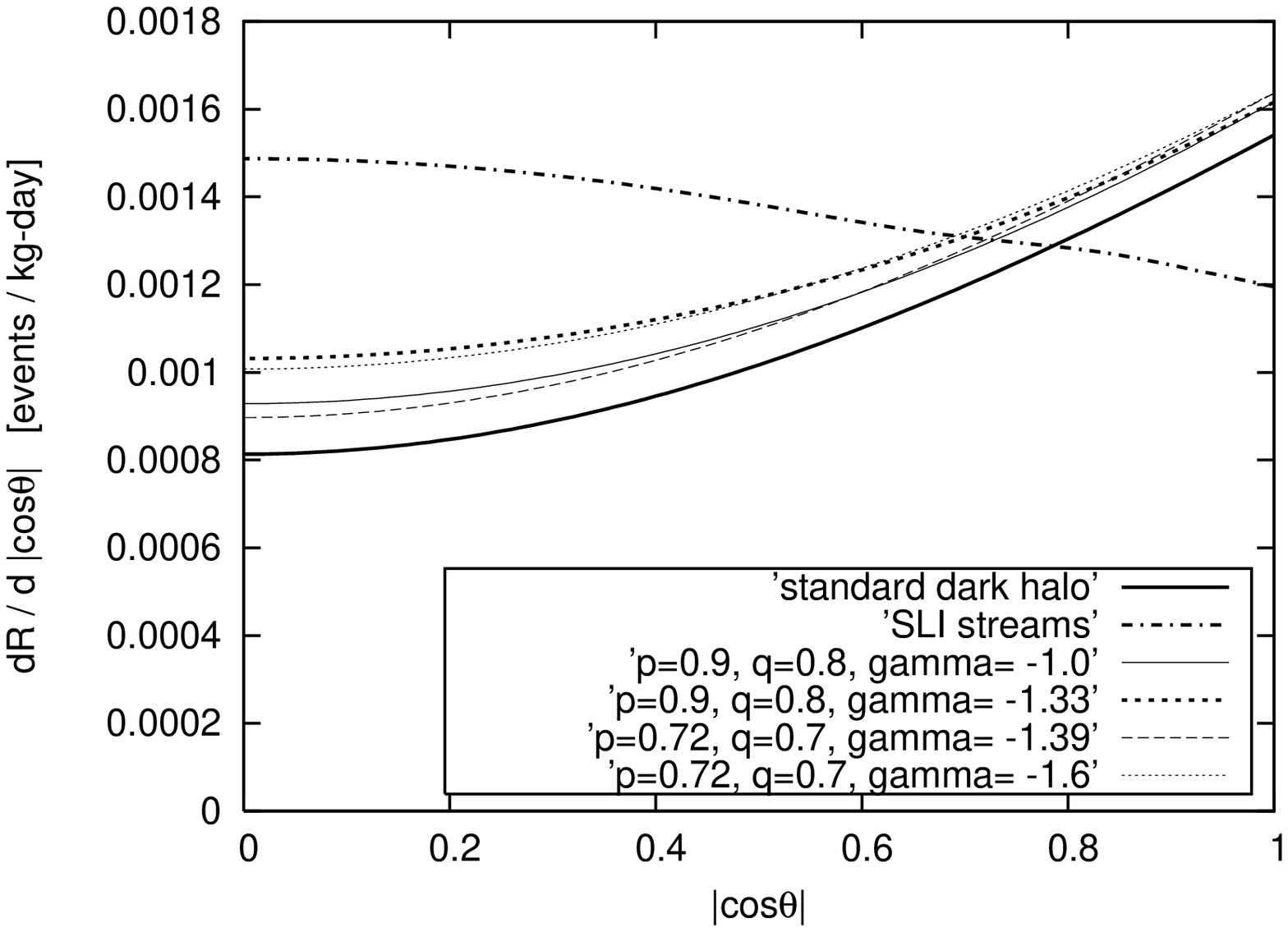}
\caption {The ``folded'' directional recoil rate  ${dR}/{d|\!\cos\theta|}$ off a CS$_{2}$ target for the standard dark halo (thick solid line), SLI streams (thick dashed-dotted line), and four logarithmic-ellipsoidal anisotropic models in which the Solar System is on the major axis of the density ellipsoid (solid, thick dotted, dashed, and dotted lines, with parameters given in the legend). In all cases, the reference direction is opposite to the direction of the Galactic rotation, $(\lambda_{\rm axis},\beta_{\rm axis})=(167\fdg340,-59\fdg574)$.}
\label{absCosSLIstdAnisoMaj}
\end{figure}

\begin{figure}[t]
\centering
\includegraphics[width=\textwidth]{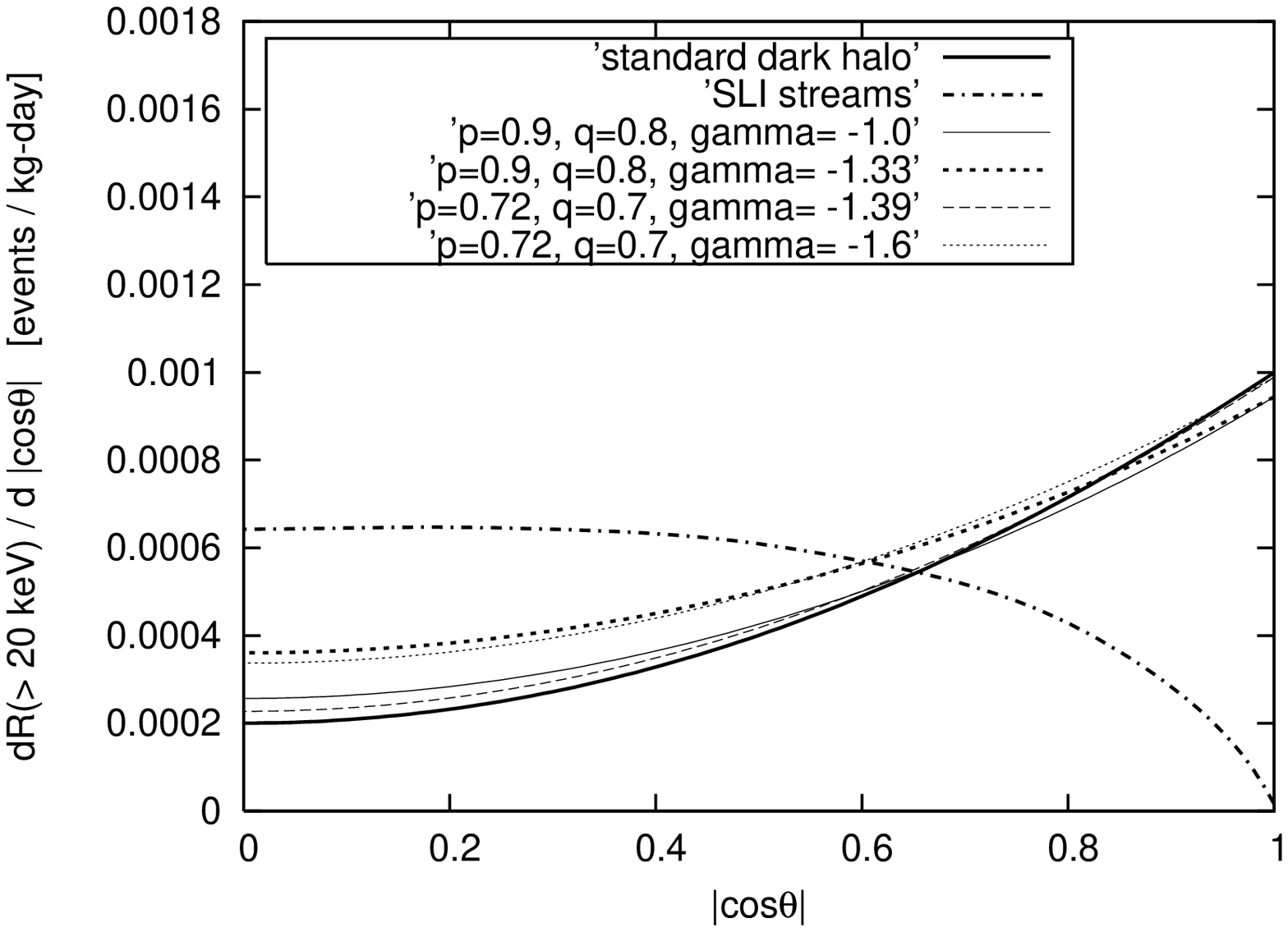}
\caption {Same as Fig.~\ref{absCosSLIstdAnisoMaj} but using a detector with threshold energy equal to 20 keV.}
\label{absCosSLIstdAnisoMajTh}
\end{figure}

We use Eq.~(\ref{dRdcosabs}) and the numerical method, Eq.~(\ref{dRdcostheRK}), for the standard dark halo (Sec.~\ref{Standard dark halo}), SLI streams (Sec.~\ref{Results for Sikivie's late-infall halo model}), and logarithmic-ellipsoidal anisotropic models (Sec.~\ref{Results for anisotropic velocity dispersion}). The results are shown in Figs.~\ref{absCosSLIstdAnisoInter}--\ref{absCosSLIstdAnisoMajTh} for the CS$_{2}$ target nuclei, where for all cases the reference direction is opposite to the direction of the Galactic rotation, $(\lambda_{\rm axis},\beta_{\rm axis})=(167\fdg340,-59\fdg574)$. In all of these figures, we show the rates for the standard dark halo and the SLI model. For the anisotropic models, the case in which the Solar System is assumed to be on the minor axis of the density ellipsoid is shown in Fig.~\ref{absCosSLIstdAnisoInter} (for an ideal zero-threshold detector) and~\ref{absCosSLIstdAnisoInterTh} (for a 20-keV threshold detector). Figs.~\ref{absCosSLIstdAnisoMaj} and~\ref{absCosSLIstdAnisoMajTh} show the analogous cases for the Solar System on the major axis of the density ellipsoid.

\begin{figure}[t]
\centering
\includegraphics[width=\textwidth]{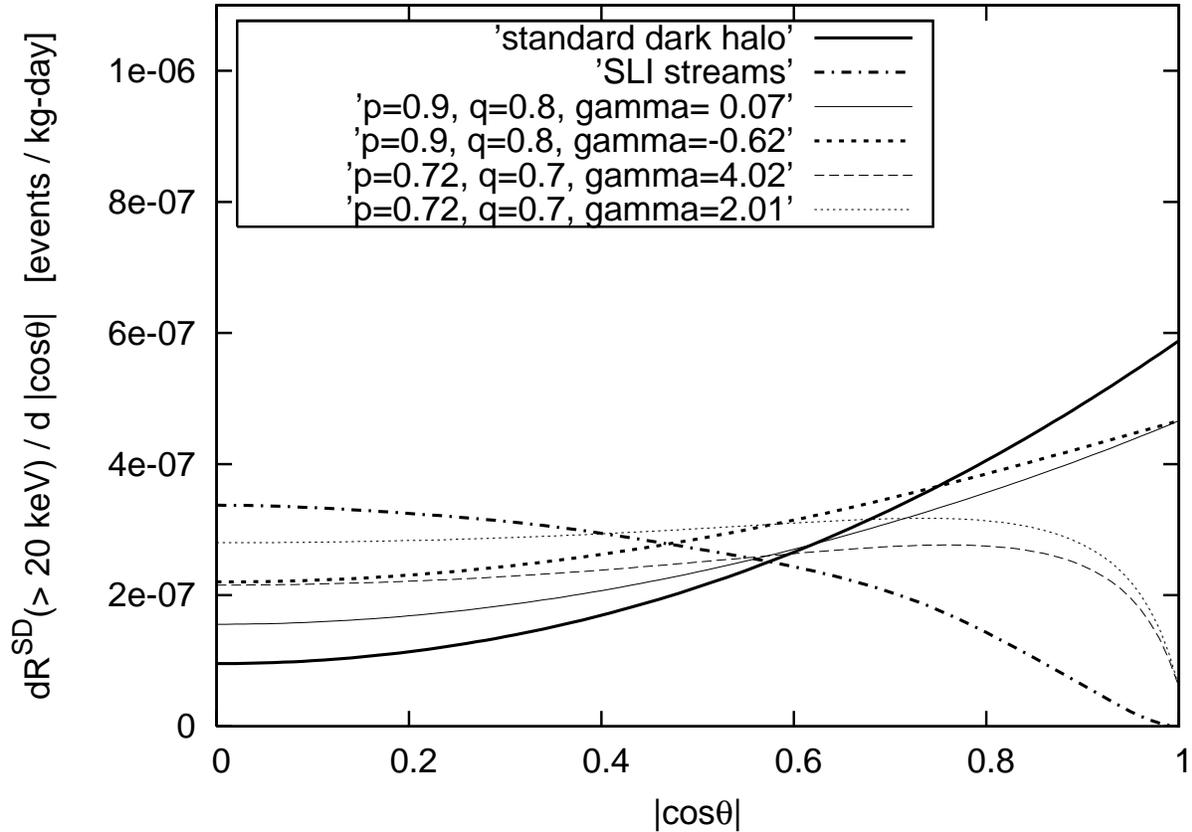}
\caption {The spin-dependent ``folded'' directional recoil rate ${dR^{\,SD}}/{d|\!\cos\theta|}$ off a CF$_{4}$ target using a detector with threshold energy equal to 20 keV, for the standard dark halo (thick solid line), SLI streams (thick dashed-dotted line), and four logarithmic-ellipsoidal anisotropic models in which the Solar System is on the minor axis of the density ellipsoid (solid, thick dotted, dashed, and dotted lines, with parameters given in the legend). In all cases, the reference direction is opposite to the direction of the Galactic rotation, $(\lambda_{\rm axis},\beta_{\rm axis})=(167\fdg340,-59\fdg574)$.}
\label{absCosSLIstdAnisoInterThSD}
\end{figure}

\begin{figure}[t]
\centering
\includegraphics[width=\textwidth]{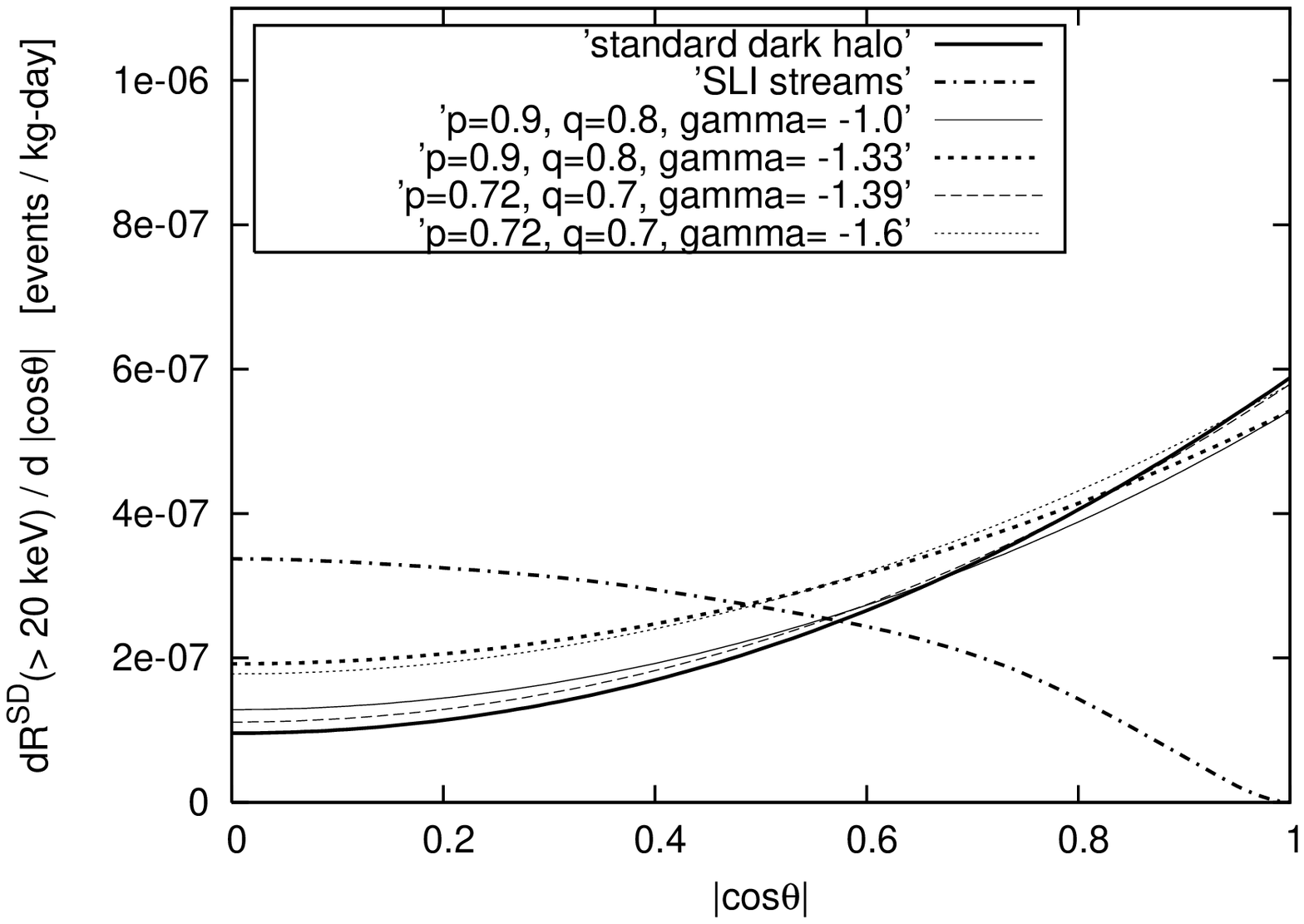}
\caption {Same as Fig.~\ref{absCosSLIstdAnisoInterThSD} but for logarithmic-ellipsoidal models in which the Solar System is on the major axis of the density ellipsoid.}
\label{absCosSLIstdAnisoMajThSD}
\end{figure}

The spin-dependent case for a 20-keV threshold detector is shown in Fig.~\ref{absCosSLIstdAnisoInterThSD} (when the Solar System is assumed to be on the minor axis of the density ellipsoid) and in Fig.~\ref{absCosSLIstdAnisoMajThSD} (when the Solar System is assumed to be on the major axis of the density ellipsoid). The shape of the $dR^{\,SD}(\!>\!{\rm 20\,keV})/d|\!\cos\theta|$ curves in Figs.~\ref{absCosSLIstdAnisoInterThSD} and~\ref{absCosSLIstdAnisoMajThSD} differs from the shape of the ${dR(\!>\!{\rm 20\,keV})}/{d|\!\cos\theta|}$ curves in Figs.~\ref{absCosSLIstdAnisoInterTh} and~\ref{absCosSLIstdAnisoMajTh} respectively because of the different masses of sulfur (S) and fluorine (F).

The $dR/d|\!\cos\theta|$ curves that show small variation with $|\!\cos\theta|$ will be harder to differentiate from an isotropic background. To make this statement quantitative, we compute the effective exposure $\Eexp$  required to distinguish a directional signal of WIMPs from a distribution uniform in $|\!\cos\theta|$ that may be due to background events. For this purpose, we use the Kolmogorov-Smirnov test, which tests if data are drawn from a given distribution (in this case, the uniform distribution). Neglecting the background, we determine the required effective exposure $\Eexp$ as follows. We use a model $dR/d|\!\cos\theta|$ to Monte-Carlo generate a $|\!\cos\theta|$ distribution of $n$ events representing the outcome of a simulated experiment with zero background. We do this for 10,000 simulated experiments. For each experiment, we compute the Kolmogorov-Smirnov statistic $D_n$, which is the maximum vertical distance between the cumulative distributions of a uniform variate and of the simulated $|\!\cos\theta|$ values.
We declare that the $|\!\cos\theta|$ distribution is non-uniform at the 90\% level of significance (i.e.\ that the experiment under question detects a WIMP signal at the 90\% significance level) when $D_n$ is greater than the upper 10\% quantile $D_{n,10}$ of the $D_n$ distribution evaluated under the null hypothesis of a uniform variate. To evaluate the probability distribution of $D_n$ under the null hypothesis, we use formulas in Refs.~\cite{Du,Steph} as implemented numerically in Refs.~\cite{NR,Pomer}. We determine the fraction of the 10,000 simulated experiments with a positive detection at the 90\% significance level. We finally increase the number $n$ of events in each simulated experiment, until the fraction of experiments with a positive detection reaches 90\%. This gives us the minimum number of events $N_e$ required to distinguish each of the halo models considered from a flat distribution at 90\% significance level in 90\% of the simulated experiments. Finally, the corresponding effective exposure $\Eexp$ is obtained by dividing $N_e$ by the integrated expected rate  above threshold $\int d|\!\cos\theta| \, [dR(\!>\!E_{\rm thr})/d|\!\cos\theta|]$ (or the total rate $\int d|\!\cos\theta| \, [dR/d|\!\cos\theta|]$ for an ideal zero-threshold detector).

\begin{table}[t]
\caption{Number of recoil events $N_{e}$ required to distinguish each of the halo models considered from a flat distribution at 90\% significance level in 90\% of the simulated experiments  using a Kolmogorov-Smirnov statistic, for a CS$_2$ zero-threshold detector and a CS$_2$ detector with threshold energy equal to 20 keV. Here $\sigma_{\rm p} = 10^{-44} {\rm cm}^2$. Also shown are the effective exposures $\Eexp$ required for each case.}
\centering
\begin{tabular}{lrrrr}
\hline\hline \\[-5ex] 
& zero threshold energy & threshold energy=20 keV  \\ [-1.5ex]
Model & \cline {1-3}
& $N_{e}$ \,\, $\Eexp$ (kg-yr) & $N_{e}$ \,\, $\Eexp$ (kg-yr) \\ [0ex]
\hline \\ [-5ex]
$\rm{standard\,\,dark\,\,halo}$ & 260 \,\,\,\,\,\,\,\,\,\,\,\,\,\,\,\,\,\,\, 330 & 39 \,\,\,\,\,\,\,\,\,\,\,\,\,\,\,\,\,\, 113 \\ [0ex]
$\rm{SLI\,\,streams}$ & 1606 \,\,\,\,\,\,\,\,\,\,\,\,\,\,\,\, 1596 & 115 \,\,\,\,\,\,\,\,\,\,\,\,\,\,\,\,\,\, 295 \\ [0ex]
\hline \\[-5ex] 
\multicolumn{3}{c}{Anisotropic\,\,logarithmic-ellipsoidal\,\,models}\,\,($\rm{Solar\,\,System\,\,is\,\,on\,\,the\,\,minor\,\,axis\,\,of\,\,the\,\,density\,\,ellipsoid}$): & \\
$p=0.9, q=0.8, \gamma= 0.07$ & 503 \,\,\,\,\,\,\,\,\,\,\,\,\,\,\,\,\,\,\, 585 & 98 \,\,\,\,\,\,\,\,\,\,\,\,\,\,\,\,\,\, 275 \\ [0ex]

$p=0.9, q=0.8, \gamma=-0.62$ & 876 \,\,\,\,\,\,\,\,\,\,\,\,\,\,\,\,\,\,\, 969 & 196 \,\,\,\,\,\,\,\,\,\,\,\,\,\,\,\,\,\, 482 \\ [0ex]

$p=0.72, q=0.7, \gamma=4.02$ & 1005 \,\,\,\,\,\,\,\,\,\,\,\,\,\,\,\, 1149 & 1199 \,\,\,\,\,\,\,\,\,\,\,\,\,\,\, 3548 \\ [0ex]

$p=0.72, q=0.7, \gamma=2.01$ & 2088 \,\,\,\,\,\,\,\,\,\,\,\,\,\,\,\, 2276 & 7500 \,\,\,\,\,\,\,\,\,\,\,\, 19116 \\ [0ex]
\hline \\[-5ex]
\multicolumn{3}{c}{Anisotropic\,\,logarithmic-ellipsoidal\,\,models}\,\,($\rm{Solar\,\,System\,\,is\,\,on\,\,the\,\,major\,\,axis\,\,of\,\,the\,\,density\,\,ellipsoid}$): & \\
$p=0.9, q=0.8, \gamma= -1.0$ & 365 \,\,\,\,\,\,\,\,\,\,\,\,\,\,\,\,\,\,\, 427 & 60 \,\,\,\,\,\,\,\,\,\,\,\,\,\,\,\,\,\, 168 \\ [0ex]

$p=0.9, q=0.8, \gamma= -1.33$ & 600 \,\,\,\,\,\,\,\,\,\,\,\,\,\,\,\,\,\,\, 668 & 115 \,\,\,\,\,\,\,\,\,\,\,\,\,\,\,\,\,\, 283 \\ [0ex]

$p=0.72, q=0.7, \gamma= -1.39$ & 301 \,\,\,\,\,\,\,\,\,\,\,\,\,\,\,\,\,\,\, 354 & 46 \,\,\,\,\,\,\,\,\,\,\,\,\,\,\,\,\,\, 130 \\ [0ex]

$p=0.72, q=0.7, \gamma= -1.6$ & 497 \,\,\,\,\,\,\,\,\,\,\,\,\,\,\,\,\,\,\, 555 & 90 \,\,\,\,\,\,\,\,\,\,\,\,\,\,\,\,\,\, 221 \\ [0ex]

\hline\hline
\end{tabular}
\label{table:Ne}
\end{table}

Table \ref{table:Ne} shows the resulting $N_e$ and $\Eexp$ for a zero threshold detector and a detector with threshold energy equal to 20 keV, assuming a CS$_{2}$ target, a WIMP mass $m=60$ GeV and a WIMP-proton cross section $\sigma_{\rm p}=10^{-44}$ cm$^2$. 

The number of events $N_e$ is independent of the value of the WIMP-proton cross section $\sigma_{\rm p}$, because it depends only on the shape, and not the height, of the $dR/d|\!\cos\theta|$ distribution. The corresponding exposure however scales as the inverse of $\sigma_{\rm p}$. Regarding the dependence on the WIMP mass, in general $N_e$ and $\Eexp$ have a complicated dependence due to the relation between the threshold energy, the target nucleus mass, and the WIMP mass. 

From Table \ref{table:Ne}, we notice the following. Even in the most favorable case of negligible background, the most optimistic case is the standard dark halo where, for a detector with a 20-keV threshold, only 39 recoil events or an effective exposure of 113 kg-yr are needed in order to distinguish the standard dark halo from a flat distribution at 90\% significance level in 90\% of the simulated experiments. For a zero threshold detector, the required number of recoil events $N_{e}$ increases to 260 events (330 kg-yr of effective exposure). These numbers are almost the same as those obtained by Morgan, Green, and Spooner \cite{MGS} for the same standard dark halo, although they used the average of $|\!\cos\theta|$ while we use the full $|\!\cos\theta|$ distribution. 

In the case of SLI streams, the required number of events $N_{e}$ is 115 (295 kg-yr of effective exposure) for a detector with 20-keV threshold, and $N_{e}$=1606 (1596 kg-yr of effective exposure) for a zero threshold detector. Distinguishing an SLI streams signal  from an isotropic background requires 3 to 5 times larger exposures than for a standard dark halo.

The anisotropic models we considered are, with few exceptions, intermediate between the standard dark halo and the SLI streams cases. The hardest cases to distinguish from a flat distribution are the two anisotropic models with parameters $p=0.72, q=0.7, \gamma=4.02$ and $p=0.72, q=0.7, \gamma=2.01$. They require effective exposures of thousands to ten thousands of kg-yr. This is due to the peculiar behavior of their ${dR(\!>\!{\rm 20\,keV})}/{d|\!\cos\theta|}$ distributions. As seen in Fig.~\ref{absCosSLIstdAnisoInterTh}, these distributions (dashed and dotted lines) exhibit a sudden drop in ${dR(\!>\!{\rm 20\,keV})}/{d|\!\cos\theta|}$ in the folded forward+backward direction $|\cos\theta|=1$. Notice that these two models are the most anisotropic models among the eight anisotropic logarithmic-ellipsoidal models considered in this study ($\gamma=4.02$ and $\gamma=2.01$, respectively). This might be the reason for such behavior.

To illustrate a spin-dependent case, we assume a CF$_4$ target, a proton-odd approximation for the F spin, a spin-dependent cross section off protons $\sigma_{\rm p}^{SD}=10^{-44} {\rm cm}^2$, and a vanishing spin-independent cross section. We examine both
 a zero threshold detector and a detector with threshold energy equal to 20 keV. Because the spin-dependent rates do not increase as $A^2$, the effective exposures required for spin-dependent interactions are longer than for the spin-independent case (at the same WIMP-proton cross section). For CF$_4$ we find them about 1,000 times longer than for CS$_2$.  For example, in the case of the standard dark halo we find that $\Eexp$ is about 4.8$\times$10$^{5}$ kg-yr for a zero threshold detector and it is about 1.6$\times$10$^{5}$ kg-yr for a detector with threshold energy equal to 20 keV. The effective exposures $\Eexp$ for the other halo models follow the same pattern and they are in the order of magnitude of 10$^{6}$ kg-yr. Such effective exposures are impractical.

To summarize, the ``folded'' directional recoil rate ${dR}/{d|\!\cos\theta|}$ can be helpful in recognizing the cases of the standard dark halo, SLI streams, and some not-too-anisotropic models. However, if the detector threshold is too low, or the degree of anisotropy too high, it may be difficult to recognize SLI streams and some other anisotropic models.

\section{Summary and Conclusions}
\label{Summary and Conclusions}

In this paper, we studied the directional recoil rate ${dR}/{d\!\cos\theta}$ of recoiling target nuclei struck by WIMPs in terms of the angle $\theta$ between the nucleus recoil direction ${\bf \hat w}$ and a chosen reference direction ${\bf \hat n}$ in the sky. We used the ecliptic coordinate system and imagined a CS$_2$ detector similar to DRIFT but with 3D read-out capabilities.

The directional recoil rate  ${dR}/{d\!\cos\theta}$ was computed and compared for different halo models that represent several WIMP velocity distributions: streams of WIMPs (Sec.~\ref{Stream of WIMPs}), the standard dark halo (Sec.~\ref{Standard dark halo}), Sikivie's late-infall halo model (SLI streams) (Sec.~\ref{Results for Sikivie's late-infall halo model}), and anisotropic logarithmic-ellipsoidal models (Sec.~\ref{Results for anisotropic velocity dispersion}). We repeated our analysis for a ``folded'' directional recoil rate ${dR}/{d|\!\cos\theta|}$ that incorporates the inability of some detectors to distinguish the beginning of a recoil track from its end (lack of head-tail discrimination). For all of the halo models considered, we compared ${dR}/{d|\!\cos\theta|}$ to an isotropic background, to examine the possibility of discriminating a WIMP signal from background noise.

We computed ${dR}/{d\!\cos\theta}$ both numerically and analytically (Sec.~\ref{Calculation of Recoil Momenta Distributions}). The numerical method (Sec.~\ref{numerical for general cases}) uses a fifth-order Cash-Karp Runge-Kutta method and can be applied to general (Gaussian and non-Gaussian) velocity distributions and any reference direction. The analytical method (Sec.~\ref{Analytic for fixed reference direction}) works only for Gaussian distributions with the reference direction ${\bf \hat n}$ aligned with the average WIMP velocity ${\bf V}$. The analytic formula was used to cross check the numerical calculation. Comparison of the numerical and analytic calculations gave the same results. In both the numerical and analytical methods, the recoil momentum function ${\hat f}(w, {\bf\hat w})$ used in the calculation of ${dR}/{d\!\cos\theta}$ was taken as the Radon transform of the velocity distribution function $f({\bf v})$. 

We applied the numerical method to the aforementioned WIMP halo models and presented the results by showing the directional recoil rate ${dR}/{d\!\cos\theta}$ as a function of the angle $\theta$ (see Sec.~\ref{Results}). For generic streams of WIMPs, we showed how varying the ratio $\sigma_{v}/V$ of the velocity dispersion $\sigma_{v}$ to the magnitude of the average WIMP velocity $V$ affect the directional recoil rate ${dR}/{d\!\cos\theta}$ (see Fig.~\ref{Fmvbx}). We also showed the effect of varying the reference direction ${\bf \hat n}$ or equivalently the stream velocity ${\bf V}$ (see Fig.~\ref{FmvbxLamBet}). 

Comparisons between the case of the SLI streams and the case of the standard dark halo showed that SLI streams produce a directional recoil rate that peaks in the opposite direction to the standard halo one (see Fig.~\ref{dRdcosRKSLIstd}). The case of streams with anisotropic logarithmic-ellipsoidal models resembles that of the standard dark halo, with small differences between the anisotropic models due to different values of their axial ratios $p$ and $q$ and anisotropy parameter $\gamma$ (see Figs.~\ref{dRdcosRKanisoInter} and ~\ref{dRdcosRKanisoMaj}).

We allowed for the difficulty of head-tail discrimination in WIMP direct detection experiments in Sec.~\ref{Head-tail discrimination}. There we introduced a ``folded'' directional recoil rate ${dR}/{d|\!\cos\theta|}$ suitable for direct comparison with experiments lacking head-tail discrimination. For both a zero threshold detector and a detector with 20-keV threshold energy, we calculated the number of recoil events ${N_{e}}$ and the effective exposure $\Eexp$ required to distinguish each of the halo models considered from a flat distribution (see Table \ref{table:Ne}). We found that in distinguishing a signal from an isotropic background noise, the ``folded'' directional recoil rate ${dR}/{d|\!\cos\theta|}$ may be effective for the standard dark halo and some of the anisotropic logarithmic-ellipsoidal models; it may be less effective for the SLI streams and other anisotropic models (see Figs.~\ref{absCosSLIstdAnisoInter}--\ref{absCosSLIstdAnisoMajTh}). In most cases, for $m=60$ GeV and $\sigma_{\rm p}=10^{-44}$ cm$^2$, exposures from few dozens to few hundreds of kg-yr of CS$_2$ would be needed to utilize the ``folded'' directional recoil rate ${dR}/{d|\!\cos\theta|}$ for the purpose of discriminating a directional WIMP signal from an isotropic background noise.

\acknowledgments

We thank Junya Kasahara for translating Japanese text for us. The research described in this paper has been supported by a graduate student scholarship from the Ministry of Higher Education in Saudi Arabia, and by the National Science Foundation through grant PHY-0456825 at the University of Utah.


\begin{thebibliography}{9}


\bibitem{LambdaCDM} http://lambda.gsfc.nasa.gov/product/map/dr2/params/lcdm\_all.cfm

\bibitem{BD} J. Binney and W. Dehnen, Mon.\ Not.\ R. Astron.\ Soc.\ {\bf 287}, L5 (1997).

\bibitem{DRIFT1} C.J. Martoff {\it et al.}, in {\it The Identification of Dark Matter} edited by N.J.C. Spooner (World Scientific, Singapore, 1996), pp. 324.

\bibitem{DRIFT2} D.P. Snowden-Ifft, C.J. Martoff, and J.M. Burwell, Phys.\ Rev.\ {\bf D 61}, 101301 (2000).

\bibitem{DRIFT3} K. Buckland, M.J. Lehner, G.E. Masek, and M. Mojaver, Phys.\ Rev.\ Lett.\ {\bf 73}, 1067 (1994).

\bibitem{DRIFT4} T. Ohnuki, D.P. Snowden-Ifft, and C.J. Martoff, Nuclear Instruments and Methods in Physics Research {\bf A463}, 142 (2001).

\bibitem{DRIFT5} C.J. Martoff {\it et al.}, Nuclear Instruments and Methods in Physics Research {\bf A440}, 355 (2000).

\bibitem{NEWAGE1} K. Miuchi {\it et al.}, arXiv:0708.2579 [astro-ph] (2007).

\bibitem{NEWAGE2} T. Tanimori {\it et al.}, Phys.\ Lett.\ {\bf B578}, 241 (2004).

\bibitem{Directional1} H. Sekiya {\it et al.}, Phys.\ Lett.\ {\bf B571}, 132 (2003).

\bibitem{Directional2} H. Sekiya {\it et al.}, in {\it Proceedings of the 5th International Workshop on The Identification of Dark Matter, Edinburgh, 2004}, edited by N.J.C. Spooner and V. Kudryavtsev (World Scientific, Hackensack, 2005), p. 378.

\bibitem{Spergel} D.N Spergel, Phys.\ Rev.\ {\bf D 37}, 1353 (1988).

\bibitem{Martoff} C.J. Martoff {\it et al.}, Phys.\ Rev.\ Lett.\ {\bf 76}, 4882 (1996).

\bibitem{DRIFTI} G.J. Alner {\it et al.}, Nuclear Instruments and Methods in Physics Research {\bf A535}, 644 (2004).

\bibitem{DRIFTII} G.J. Alner {\it et al.}, Nuclear Instruments and Methods in Physics Research {\bf A555}, 173 (2005).

\bibitem{DRIFTIII} J.C. Davies {\it et al.}, in {\it Proceedings of the 5th International Workshop on The Identification of Dark Matter, Edinburgh, 2004}, edited by N.J.C. Spooner and V. Kudryavtsev (World Scientific, Hackensack, 2005), p. 242.

\bibitem{CHK} C.J. Copi, J. Heo, and L.M. Krauss, Phys.\ Lett.\ {\bf B461}, 43 (1999).

\bibitem{CK} C.J. Copi and L.M. Krauss, Phys.\ Rev.\ {\bf D 63}, 043507 (2001).

\bibitem{SavGonFre} C. Savage, P. Gondolo, and K. Freese, Phys.\ Rev.\ {\bf D 70}, 123513 (2004).

\bibitem{Engel} J. Engel, Phys.\ Lett.\ {\bf B264}, 114 (1991).

\bibitem{EllisFlores} J. Ellis and R.A. Flores, Phys.\ Lett.\ {\bf B263}, 259 (1991).

\bibitem{LewinSmith} J.D. Lewin and P.F. Smith, Astropart. Phys. {\bf 6}, 87 (1996).

\bibitem{Gon} P. Gondolo, Phys.\ Rev.\ {\bf D 66}, 103513 (2002).

\bibitem{Radon} J. Radon, Ber. Verh. S\"{a}chs. Akad. Wiss. Leipzig. Math. Phys. K1. {\bf 69}, 262 (1917).

\bibitem{FGN} K. Freese, P. Gondolo, and H.J. Newberg, Phys.\ Rev.\ {\bf D 71}, 043516 (2005).

\bibitem{MGS} B. Morgan, A.M. Green, and N.J.C. Spooner, Phys.\ Rev.\ {\bf D 71}, 103507 (2005).

\bibitem{MorganGreen} B. Morgan and A. M. Green, Phys.\ Rev.\ {\bf D 72}, 123501 (2005).

\bibitem{GreenMorgan} A. M. Green and B. Morgan, Astropart. Phys. {\bf 27}, 142 (2007).

\bibitem{HostHansen} O. Host and S.H. Hansen, JCAP {\bf 06}, 016 (2007).

\bibitem{Gas} J. Gascon, arXiv:astro-ph/0504241.

\bibitem{Lehner} M.J. Lehner {\it et al.}, arXiv:astro-ph/9905074.

\bibitem{AG} M.S. Alenazi and P. Gondolo, Phys.\ Rev.\ {\bf D 74}, 083518 (2006).

\bibitem{NR} W.H. Press {\it et al.}, {\it Numerical Recipes in C : The Art of Scientific Computing}, 2nd ed. (Cambridge Univ.\ Press, 1992).

\bibitem{STW1} P. Sikivie, I.I. Tkachev, and Y. Wang, Phys.\ Rev.\ Lett.\ {\bf 75}, 2911 (1995).

\bibitem{STW2} P. Sikivie, I.I. Tkachev, and Y. Wang, Phys.\ Rev.\ {\bf D 56}, 1863 (1997).

\bibitem{Si} P. Sikivie, Phys.\ Rev.\ {\bf D 60}, 063501 (1999).

\bibitem{Sikivie} P. Sikivie, Nucl.\ Phys.\ B (Proc.\ Suppl.) {\bf 72}, 110 (1999).

\bibitem{GG} G. Gelmini and P. Gondolo, Phys.\ Rev.\ {\bf D 64}, 023504 (2001).

\bibitem{WJ} W. Dehnen and J.J. Binney, Mon.\ Not.\ R. Astron.\ Soc.\ {\bf 298}, 387 (1998).

\bibitem{Mooreetal} B. Moore {\it et al.}, Phys.\ Rev.\ {\bf D 64}, 063508 (2001).

\bibitem{HWS} A. Helmi, S.D.M. White, and V. Springel, Phys.\ Rev.\ {\bf D 66}, 063502 (2002).

\bibitem{Green} A.M. Green, Phys.\ Rev.\ {\bf D 66}, 083003 (2002).

\bibitem{ECdeZ} N.W. Evans, C.M. Carollo, and P.T. de Zeeuw, Mon.\ Not.\ R. Astron.\ Soc.\ {\bf 318}, 1131 (2000).

\bibitem{Du} J. Durbin, The Annals of Mathematical Statistics {\bf 39}, 398 (1968).

\bibitem{Steph} M.A. Stephens, J. R. Stat. Soc. Ser. B Methodol. {\bf 32}, 115 (1970).

\bibitem{Pomer} J. Pomeranz, Communications of the ACM {\bf 17}, 703 (1974).

\end{thebibliography}
\end{document}